\documentclass[onefignum,onetabnum]{siamonline181217}

\usepackage{amsfonts}
\usepackage{mathtools}
\usepackage{hyperref}
\usepackage{autonum}
\usepackage{graphicx}
\usepackage{subcaption}
\usepackage{moreverb}

\usepackage{pgfplots}
\usepackage{pgfplotstable}
\usepgfplotslibrary{colorbrewer}
\usepgfplotslibrary{groupplots}
\pgfplotsset{compat=newest}
\usepackage{tikz}
\usetikzlibrary{external}
\allowdisplaybreaks

\newcommand{\x}{\vct{x}}
\newcommand{\y}{\vct{y}}

\newcommand{\Id}{\operatorname*{Id}}
\newcommand{\sign}{\operatorname*{sign}}

\newcommand{\norm}[1]{\lVert #1 \rVert}
\newcommand{\abs}[1]{| #1 |}

\newcommand{\inv}{^{\scalebox{0.7}[1.0]{\( - \)}1}}
\newcommand{\tp}{^t}
\renewcommand{\exp}[1]{e^{#1}}
\newcommand{\Exp}[1]{\operatorname*{Exp}\left(#1\right)}
\renewcommand{\i}{\mathrm{i}}
\renewcommand{\d}{\operatorname{d}\!}
\newcommand{\tr}{\operatorname{tr}}

\newcommand{\mat}[1]{\begin{bmatrix} #1 \end{bmatrix}}
\newcommand{\vct}[1]{\pmb{#1}}
\newcommand{\tns}[1]{\pmb{#1}}
\newcommand{\tnsfour}[1]{\pmb{#1}}

\newcommand{\bbc}[1]{\left( #1 \right)}
\newcommand{\rbc}[1]{\left[ #1 \right]}

\renewcommand{\div}{\nabla\cdot}
\newcommand{\grad}{\nabla}

\newcommand{\Fi}{\mathcal{F}^{-1}} 

\newcommand{\R}{\mathbb{R}} 
\newcommand{\N}{\mathbb{N}} 

\renewcommand{\H}{\mathcal{H}} 

\newcommand{\E}[1]{\operatorname{\mathbb{E}}\rbc{#1}}
\newcommand{\Var}[1]{\operatorname{Var}\rbc{#1}}
\newcommand{\Gaussian}{\mathcal{N}}

\newcommand{\WN}{\mathcal{W}} 
\newcommand{\WNhat}{\hat{\mathcal{W}}} 

\newcommand{\BesselK}[1]{\mathcal{K}_{#1}}
\newcommand{\erf}{\operatorname{erf}}

\newcounter{rownumber}

\newcommand{\Phase}{\chi} 	
\newcommand{\Int}{m} 		
\newcommand{\cov}{\mathcal{C}}
\newcommand{\vf}{\phi}
\newcommand{\vfm}{\vf_0}
\newcommand{\corrlen}{\ell}

\newcommand{\Matern}[1]{\mathcal{M}_{#1}} 

\newcommand{\Shape}{\tns{\Theta}}

\newcommand{\Vtot}{V_{total}}
\newcommand{\Vpore}{V_{pore}}
\newcommand{\Spore}{S_{pore}}
\newcommand{\Sphericity}{\operatorname{Sphericity}}
\newcommand{\RelSize}{r}

\newcommand{\prmspace}{\Xi}
\newcommand{\prm}{\xi}
\newcommand{\prms}{\vct{\prm}}
\newcommand{\Likelihood}{\operatorname{Likelihood}}

\newcommand{\disp}{\vct{u}}
\newcommand{\strain}{\tns{\varepsilon}}
\newcommand{\stress}{\tns{\sigma}}
\newcommand{\Macrostrain}{\tns{E}}
\newcommand{\Macrostress}{\tns{\Sigma}}
\newcommand{\dv}{\tns{s}}
\newcommand{\nbhd}{O}

\newcommand{\bulk}{K}
\newcommand{\shear}{G}
\newcommand{\Stiffness}{\tnsfour{C}}
\newcommand{\QoI}{Q}

\newsiamremark{rmk}{Remark}

\pgfplotsset{axis standard/.style={
		width=0.49\textwidth,
		grid=major,
		y tick label style={font=\scriptsize,}, 
		x tick label style={font=\scriptsize, /pgf/number format/.cd, fixed, precision=2},
		/pgf/number format/.cd, fixed, precision=2,
		y label style={at={(-0.15,0.5)}},
		legend cell align={left},
		legend style={at={(0.5,-0.1)},anchor=north,font=\scriptsize, draw=none},
	}
}

\pgfplotsset{nu ticks/.style={xtick={1, 3, 5, 7, 9},}}
\pgfplotsset{enable error bars/.style={error bars/.cd, y dir=both, y explicit, error bar style={thin},},}
	
\pgfplotsset{line standard/.style={
		thick,
		smooth,
		mark=*,
		mark options={solid},
		mark size=1pt
	}
}

\pgfplotsset{select coords between index/.style 2 args={
		x filter/.code={
			\ifnum\coordindex<#1\fi
			\ifnum\coordindex>#2\fi
		}
}}

\ifpdf
\hypersetup{
	pdftitle={A~statistical~framework~for~generating~microstructures of~two-phase~random~materials: application~to~fatigue~analysis},
	pdfauthor={U.~Khristenko, A.~Constantinescu, P.~Le~Tallec, J.~T.~Oden and B.~Wohlmuth}
}
\fi

\headers{A statistical framework for generating microstructures}{U.~Khristenko, A.~Constantinescu, P.~Le~Tallec, J.~T.~Oden and B.~Wohlmuth}

\title{A~statistical~framework~for~generating~microstructures of~two-phase~random~materials: application~to~fatigue~analysis\thanks{
		Submitted to the editors May 1, 2019.
		\funding{
			The support by the DFG under grant WO671/11-1 and WO671/15-2 (within the Priority Programme SPP 1748) is gratefully acknowledged by BW and UK. 
			The work of JTO was supported by the US Department of Energy, Office of Science, Office of Advanced Scientific Computing Research (MMI CCS), under award DE-SG00/9393.
			LMS work has been partly supported by the Andr\'e Citro\"en Chair, and the part of the work has been finalized during the stay of PLT at Oden Institute under the JTO Visitor's program. 
		}
	}}

\author{Ustim~Khristenko\thanks{Technical University of Munich, Germany, Department of Mathematics, Chair~of~Numerical~Mathematics~(M2)
		(\email{khristen@ma.tum.de}, \email{wohlmuth@ma.tum.de}).}
	\and Andrei~Constantinescu\thanks{Laboratoire de M{\'e}canique des Solides (LMS), {\'E}cole Polytechnique, Palaiseau, France
		(\email{andrei.constantinescu@polytechnique.edu}, \email{patrick.letallec@polytechnique.edu}).}
	\and Patrick~Le~Tallec\footnotemark[3]
	\and J.~Tinsley~Oden\thanks{Oden Institute for Computational Engineering and Sciences, The~University~of~Texas~at~Austin 
		(\email{oden@ices.utexas.edu}).}
	\and Barbara~Wohlmuth\footnotemark[2]}

\begin{document}
	
	\maketitle
	
	\begin{abstract}
		Random microstructures of heterogeneous materials play a crucial role in the material macroscopic behavior and in predictions of its effective properties.
		A common approach to modeling random multiphase materials is to develop so-called surrogate models approximating statistical features of the material.
		However, the surrogate models used in fatigue analysis usually employ simple microstructure, consisting of ideal geometries such as ellipsoidal inclusions, which generally does not capture complex geometries.			
		In this paper, we introduce a simple but flexible surrogate microstructure model for two-phase materials through a level-cut of a Gaussian random field with covariance of Mat\'ern class.
		Such parametrization of the covariance function allows for the representation of a few key design parameters while representing the geometry of inclusions in a more general setting for a large class of random heterogeneous two-phase media.	
		In addition to the traditional morphology descriptors such as porosity, size and aspect ratio, it provides control of the regularity of the inclusions interface and sphericity.
		These parameters are estimated from a small number of real material images using Bayesian inversion.
		An efficient process of evaluating the samples, based on the Fast Fourier Transform, makes possible the use of Monte-Carlo methods to estimate statistical properties for the quantities of interest in a given material class.
		We demonstrate the overall framework of the use of the surrogate material model in application to the uncertainty quantification in fatigue analysis, its feasibility and efficiency, and its role in the microstructure design.
	\end{abstract}
	



	\section{Introduction}

Real materials, employed in engineering applications or that occur in nature, possess random heterogeneous microstructures, which play a crucial role in their macroscopic behavior and in determining their effective properties~\cite{hill1965self,hill1967,castaneda1991,castaneda1995effect,hashin1963}.
The determination of how particular microstructures lead to effective macroscale properties has been a major goal of material science for decades.
The particular features of the microstructure of a material, as for example porosities, cracks, polycrystalline shape and texture, etc., have an important impact on the lifetime of an industrial structure and highly depend on the choice of manufacturing process.
The standard paradigm is to split the design analysis of structures in two steps.
First, a global mechanical analysis is performed on the complete structure using the constitutive law of the material, and second, a local fatigue analysis based on the global stabilized stress-strain response is applied to a local elementary volume element of material.
Standard fatigue analysis is generally based on a phenomenological formulae and only characterizes the average response of the material, neglecting the microstructure variability at the material.
A micromechanical model representing the microstructure as a plastic inclusion has been initially proposed by Orowan~\cite{orowan1949fracture} and interpreted and developed later for polycrysalline solids in~\cite{dang1993macro}.
Recently, methods have been proposed to analyze the statistical output of the microstructure and to estimate not only the average statistical lifetime, but also the spread of the lifetime due to the variability of the microstructure~\cite{guerchais2014micromechanical}.

A common approach to modeling random multiphase heterogeneous materials  is to develop so-called surrogate models that ideally mimic key macroscale features of the material.
These models approximate the geometric distribution of phases, and effective properties are calculated using homogenization, often implemented using Monte-Carlo methods.
Such surrogates for two-phase materials have been used to study and to characterize properties of the porous media, e.g., permeability~\cite{bryant1992prediction} or effective constitutive models~\cite{gurson1977continuum,gologanu1993approximate,monchiet2006approximate,danas2009,danas2012,madou2012}, in fatigue and damage analysis~\cite{guaruajeu2000micromechanical,mbiakop2015void}, in microstructure design~\cite{bessa2017framework}.
However, these models usually employ simple microstructure, consisting only of ideal geometries such as ellipsoidal inclusions.
Understandably, these idealized surrogate models have limited use as they generally do not capture complex geometries and distributions of phases such as inclusions.

Alternatively, realistic microstructures can be obtained experimentally by extracting and analyzing samples from real materials, e.g., Computed Tomography (CT) or Scanning Electron Microscopy (SEM).
These methods are time-consuming and expensive, possibly destructive, and cannot be deployed on a large database.
In order to complete the experimental databases, one can use algorithms to generate artificial microstructures, based on the statistical information from only a small number of original samples.
Such stochastic reconstruction methods provide effective tools for efficient reproduction of heterogeneous microstructures~\cite{rintoul1997reconstruction,yeong1998reconstructing,lu1992lineal,okabe2005pore}.
Recently developed methods based on convolutional neural networks~\cite{mosser2017reconstruction,cang2018improving} fall into this category.
Although these methods can lead to an accurate microstructure reconstruction, their mathematical and statistical structures are often based on heuristic arguments.


In this paper, we first propose a simple but flexible surrogate microstructure model for two-phase materials.
Key design parameters of the microstructure are provided, while the geometry of inclusions or pores is represented in a general setting, which permits a close representation of the microstructure of materials.
In a second stage, we use the surrogate microstructure as the representative volume element (RVE) for a fatigue analysis of the Dang Van type~\cite{dang1993macro,papadopoulos1994new,constantinescu2003unified,charkaluk2009revisiting}.
More precisely, we focus on the high cycle fatigue regime where both structure and microstructure are in an elastic shakedown regime under an external cyclic loading.
The novelty is two fold: on one hand, we introduce realistic microstructures in the analysis, and on the other hand, we propose tools for a statistical analysis of the results at the local scale of the elementary volume element.

We define our surrogate material model through a level-cut of a Gaussian random field~\cite{quiblier1984new,berk1987scattering,teubner1991level,levitz1998off,santosa2003levelset}, called the intensity field.
Such a model is entirely controlled by the mean and the covariance function of this field.
The choice of the covariance defines the morphology of the microstructure.
Thus, considering the covariances from a given class constrains our design parameter space.
In our model we consider covariances of Mat\'ern class~\cite{whittle1954stationary,matern1986spatial,handcock1993bayesian}.
In addition to the traditional morphology descriptors such as porosity, size and aspect ratio, this covariance class provides control of the smoothness of the inclusion interface.
This regularity parameter is also related to the sphericity of pores or inclusions.

While defined only by a few "design" parameters, such a surrogate model is able to mimic a large class of random heterogeneous two-phase media.
These parameters can be estimated from a small number of real material images by approximating the image statistical descriptors by the surrogate ones.
Here, we use Bayesian inversion to find the probability distributions of the design parameters, from which they are then drawn for each material sample.

The intensity field is constructed by convolving the Mat\'ern-type covariance with the white noise.
In the case of statistically homogeneous media, this can be done by Fast Fourier Transform.
Moreover, the Fourier Transform of the Mat\'ern covariance is given in a closed analytical form.
This leads to a very efficient sampling process, which makes possible the use of Monte-Carlo methods to evaluate statistical properties for the quantities of interest in a given material class.


An important goal of this work is to demonstrate the overall framework of the use of the surrogate material model in applications to homogenization and fatigue analysis, its feasibility and efficiency, and its role in the microstructure design.

The paper is structured as follows.
First, we describe in detail the properties and assumptions underlying the construction of a class of surrogate material models in Section~\ref{sec:Model}.
We present the level-set of Gaussian field model (Subsection~\ref{sec:levelset}), the Mat\'ern covariance (Subsection~\ref{sec:Matern}) and the microstructure reconstruction technique from real material images (Subsection~\ref{sec:Reconstruction}).
In Section~\ref{sec:LE}, we give the formulation of the associated linear elasticity problem together with the description of quantities of mechanical interest used in fatigue analysis.
The simulation results are taken up in Section~\ref{sec:Sim}, including analysis and prediction of damage evolution and failure (Subsection~\ref{sec:AppRealMat}), and a discussion of the use of the resulting models for the design of microstructure employing the process parameters (Subsection~\ref{sec:AppSensitivity}).
We present final discussions and conclusions in Section~\ref{sec:Conclusion}.

	\section{Surrogate material model}\label{sec:Model}

\label{sec:Surrogate}

Let us consider a random two-phase composite material in a bounded domain~$D\subset\R^d$, $d=2,3$, the phases representing a matrix with inclusions (e.g., pores, precipitates, etc.).
For a particular microstructure realization, the material phases are defined by the characteristic function
\begin{equation}\label{eq:char}
\Phase(\x; \omega) = \begin{cases}
1, & \x\text{ in inclusion}, \\
0, & \x\text{ in matrix},
\end{cases}
\qquad \x\in D\subset\R^d, \quad \omega\in\Omega,
\end{equation}
where $\Omega$ is the space of the material samples (as in~\cite{torquato2013random}), such that each sample point~$\omega\in\Omega$ corresponds to a realization of a spatial random field~$\Phase(\x; \omega)$, $\x\in D$.
Thus, for a particular realization~$\omega$, the spatial distribution of a material property~$\kappa$ over~$D$ can be written as
\begin{equation}\label{eq:kappa_chi}
\kappa(\x; \omega) = \kappa_I \cdot\Phase(\x; \omega) + \kappa_M \cdot (1-\Phase(\x; \omega)), \qquad \x\in D, \quad\omega\in\Omega,
\end{equation}
where $\kappa_I$ and $\kappa_M$ are the corresponding properties of the inclusions and the matrix, respectively.

\subsection{Level-cut of a Gaussian field}
\label{sec:levelset}

The characteristic function of the material phase can be expressed as a level-cut of an intensity field~$\Int(\x; \omega)$:
\begin{equation}\label{eq:levelcut}
\Phase(\x; \omega) = \begin{cases}
1, & \text{if } \abs{\Int(\x; \omega)} \ge \tau, \qquad\text{inclusions},\\
0, & \text{if } \abs{\Int(\x; \omega)} < \tau, \qquad\text{matrix},
\end{cases}
\end{equation}
$\omega\in\Omega$, where the level~$\tau\ge0$ controls the volume fractions of the inclusions.
We define the intensity~$\Int(\x)$ as a zero-mean Gaussian random field~\cite{adler2010geometry} with covariance~$\cov(\x,\y)$.
Suggested in earlier works of~\cite{cahn1965phase,quiblier1984new,berk1987scattering}, the level-set model is widely used in porous media reconstruction~\cite{teubner1991level,levitz1998off,ilango2013reconstruction} and in geometric inverse problems for interfaces~\cite{santosa2003levelset,dunlop2017hierarchical,chada2018parameterizations}.

Given~(\ref{eq:levelcut}), the formula~(\ref{eq:kappa_chi}) can be rewritten as
\begin{equation}\label{eq:kappa_sign}
\kappa(\x) = \frac{\kappa_I + \kappa_M}{2} + \frac{\kappa_I - \kappa_M}{2}\cdot\sign\left(\abs{\Int(\x)}-\tau\right).
\end{equation}
Let us denote the standard deviation of~$\Int(\x; \omega)$ by~$\sigma=\sqrt{\cov(\x,\x)}$.
Then, since $\sign\!\left(\abs{\Int(\x)}\!-\!\tau\right)$ $\!=\! \sign\left(\frac{\abs{\Int(\x)}}{\sigma}-\frac{\tau}{\sigma}\right)$, we can consider the Gaussian field $\sigma\inv\Int(\x)$ instead of~$\Int(\x)$ with the level parameter~$\frac{\tau}{\sigma}$.
Thus, without loss of generality, we can assume~$\cov(\x,\x)=1$ and thus $\sigma=1$.


The first two moments of the random field~$\Phase(\x,\omega)$ are given in closed form, similar to~\cite{berk1991scattering,teubner1991level}, by the following lemma (see~\ref{apx:Moments} for the proof).

\begin{lemma}\label{lem:moments}
	Let $S_1 = \E{\Phase(\x)}$ and $S_2(\vct{x},\vct{y}) = \E{\Phase(\x)\,\Phase(\y)}$ be respectively the one- and two-point correlation functions~\cite{torquato2013random} of the random field~$\Phase(\x,\omega)$ defined by~\eqref{eq:levelcut}.
	Then, they can be written as
	\begin{align}
		S_1 = \vfm &= \sqrt{\frac{2}{\pi}}\int\limits_{\tau}^{\infty}\exp{-\frac{1}{2}t^2} \d t , \label{eq:S1}\\
		S_2(\vct{x},\vct{y}) 
		&= \frac{2}{\pi} \int\limits_0^{\cov(\x,\y)}\exp{-\frac{\tau^2}{1-t^2}}\cosh\left(\frac{\tau^2 t}{1-t^2}\right) \frac{\d t}{\sqrt{1-t^2}} + \vfm^2 \\
		&= 2\vfm - 4T\left(\tau, \sqrt{\frac{1-\cov(\x,\y)}{1+\cov(\x,\y)}} \right) - 4T\left(\tau, \sqrt{\frac{1+\cov(\x,\y)}{1-\cov(\x,\y)}} \right)
		,\label{eq:S2}
	\end{align}
	where 
	$T(\tau,x)=\frac{1}{2\pi}\int\limits_{0}^{x}\exp{-\frac{\tau^2}{2}(t^2+1)} \frac{\d t}{t^2+1} $ is Owen's T function~\cite{owen1956tables,patefield2000fast}, and 
	$\vfm$ denotes the expected volume fraction of the inclusions (or porosity).
	It also holds
	\begin{equation}
		S_2(\x,\x)=\vfm, \qquad \lim_{\norm{\x-\y}\rightarrow\infty}S_2(\x,\y)=\vfm^2.
	\end{equation}
\end{lemma}

From~(\ref{eq:S1}), it follows that the level-set parameter~$\tau$ is uniquely defined through the given expected volume fraction~$\vfm$,
\begin{equation}\label{eq:tau_vf}
\tau(\vfm) = \sqrt{2}\erf^{-1}(1-\vfm),
\end{equation}
where $\erf(x)$ denotes the Gauss error function.


\begin{rmk}
	Defining in~\eqref{eq:levelcut} the phase through the absolute value of~$\Int(\x; \omega)$, see also~\cite{berk1987scattering,berk1991scattering}, allows us to distinguish the matrix and the inclusions for all values of the inclusions volume fraction~$\vfm$. 
	This departs from the model in~\cite{quiblier1984new,teubner1991level,ilango2013reconstruction}, where the phase is defined simply by the sign of $\Int(\x; \omega)-\tau$, $\tau\in\R$, and the "matrix" becomes "inclusions" when $\vfm>0.5$ (see Figure~\ref{fig:compare_models}).
	
	\begin{figure}[!ht]
		\centering
		\newcommand{\size}{0.35\textwidth}
		
		\begin{subfigure}{\textwidth}
			\centering
			\frame{\includegraphics[width=\size]{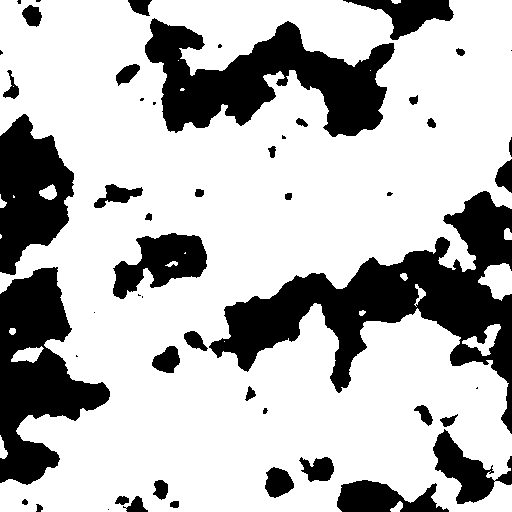}}\hspace{4ex}
			\frame{\includegraphics[width=\size]{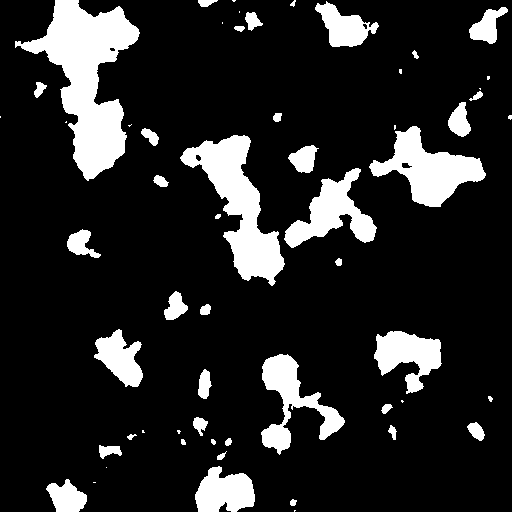}}
			\caption{Black phase defined by $\Int(\x; \omega) \ge \tau(\vfm)$ for $\vfm=0.2$ (left) and $0.8$ (right)}\label{sbfig:model0}
		\end{subfigure}
		\vspace{2ex}
		
		\begin{subfigure}{\textwidth}
			\centering
			\frame{\includegraphics[width=\size]{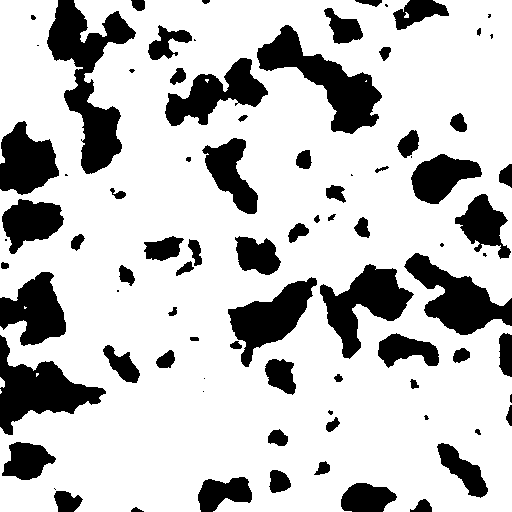}}\hspace{4ex}
			\frame{\includegraphics[width=\size]{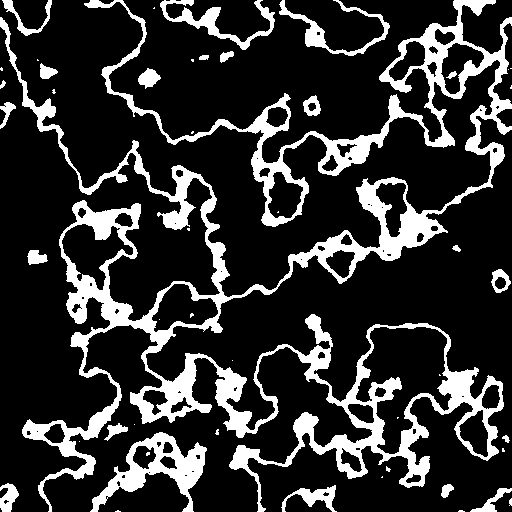}}
			\caption{Black phase defined by $\abs{\Int(\x; \omega)} \ge \tau(\vfm)$  for $\vfm=0.2$ (left) and $0.8$ (right).}\label{sbfig:model1}
		\end{subfigure}
		
		\caption{Comparison of two level-cut models for "inclusions" (black phase) defined by (\subref{sbfig:model0}) $\Int(\x; \omega) \ge \tau(\vfm)$ and (\subref{sbfig:model1}) $\abs{\Int(\x; \omega)} \ge \tau(\vfm)$, for fixed $\nu = 1.5$ and $\corrlen=0.05$, constructed from the same white noise realization. The black phase volume fraction is $\vfm=0.2$ (left) and $0.8$ (right).
			The model~(\subref{sbfig:model1}), using the absolute value, allows to distinguish the matrix and the inclusions for high volume fraction values.
		}		
		\label{fig:compare_models}
	\end{figure}
\end{rmk}


In this work, we primarily focus on statistically isotropic materials~\cite{torquato2013random}.
The isotropy constraint on~$\cov(\x,\y)$ is taken for simplicity, but it does not limit the generality of the approach. 
In this case, the covariance~$\cov(\x,\y)$ and, thus, the two-point correlation~$S_2(\x,\y)$ are rotationally invariant (stationary) and depend only on the distance between the points:
\begin{equation}
	S_2(\x,\y) = S_2(\norm{\x-\y}), \qquad \cov(\x,\y) = \cov(\norm{\x-\y}).
\end{equation}
In what follows, we will write a stationary function of two arguments $\x$ and $\y$ as a function of the distance $r = \norm{\x-\y}$.


\subsection{Mat\'ern covariance}
\label{sec:Matern}

The microstructure morphology of the above level-cut model is defined by the covariance function~$\cov(\vct{x},\vct{y})$ of the intensity field~$\Int(\x; \omega)$.
Parametrization of this covariance provides the so-called "design" parameters of the microstructure.
In general, this covariance can have an infinite dimensional parameter space.
In order to reduce the model and to have a limited number of design parameters, one can consider a projection of the material onto a reduced material class, that is, one may consider a parameterized class of covariance kernels.
Here we use the Matérn covariance class~\cite{whittle1954stationary,matern1986spatial,handcock1993bayesian}, that is of the form
\begin{equation}\label{eq:MaternCovariance}
\cov(\vct{x},\vct{y}) = \sigma^2\,\Matern{\nu}\left(\sqrt{2\nu}\cdot r\right), \qquad r = \sqrt{(\vct{x}-\vct{y})\tns{\Theta}\inv(\vct{x}-\vct{y})},
\end{equation}
with unit standard deviation~$\sigma=1$ and Mat\'ern kernel
\begin{equation}\label{eq:MaternFunction}
	\Matern{\nu}(x) = \frac{1}{2^{\nu-1}\Gamma(\nu)}\, x^\nu \,\BesselK{\nu}(x)
\end{equation}
where $\Gamma$ and $\BesselK{\nu}$ denote the Euler Gamma function and the modified Bessel function of the second kind~\cite{abramowitz1965handbook,bateman1954tables,watson1995treatise}, respectively.
The scalar parameter~$\nu>0$ defines the differentiability (smoothness) of the field, while the second order tensor~$\tns{\Theta}$ defines the shape of inclusions.
In particular, $\tns{\Theta}=\corrlen^2\,\tns{\Id}$ corresponds to an isotropic covariance with correlation length~$\corrlen$.
In general, $\tns{\Theta}$ can be written through a rotation and scaling as
\begin{equation}
	\tns{\Theta} = \tns{R}^t \mat{\corrlen_1^2 & & \\ & \ddots & \\ & & \corrlen_d^2} \tns{R},
\end{equation}
where the matrix~$\tns{R}$ defines a rotation of axes in~$\R^d$, and $\corrlen_k$, $k=1,\ldots,d$, are the correlation lengths in each axis direction.

The Mat\'ern covariance model~(\ref{eq:MaternCovariance}) is widely used in statistics~\cite{stein2012interpolation,gneiting2012studies}, geostatistics~\cite{minasny2005matern} and machine learning~\cite{williams2006gaussian}, and represents a large class of popular covariance kernels.
In particular, when $\nu$ varies from $0.5$ to infinity, it presents a continuous family of kernels from the exponential function to the square exponential (Gaussian), respectively.


It is shown in~\cite{whittle1954stationary} that the Mat{\'{e}}rn covariance function is related to the Green's function of a stochastic PDE.
Namely, the Gaussian random field~$\Int(\x;\omega)$ with Mat\'ern covariance~(\ref{eq:MaternCovariance}) is the solution of the following linear stochastic PDE in $\R^d$, $d\in\N$, \cite{lindgren2011explicit,roininen2014whittle,croci2018efficient}
\begin{equation}\label{eq:SPDE}
\left(\Id - \frac{1}{2\nu}\div (\tns{\Theta}\nabla) \right)^{\frac{1}{2}(\nu + \frac{d}{2})} \Int(\x; \omega) = \eta\,\WN(\x; \omega),
\end{equation}
\noindent where $\WN \sim \Gaussian(0, \Id)$ is a spatial white noise in~$\R^d$, and the normalization parameter~$\eta$ is given by
\begin{equation}
\eta^2 = \frac{(2\pi)^{\frac{d}{2}}\sqrt{\det\tns{\Theta}}\,\Gamma(\nu+d/2)}{\nu^{\frac{d}{2}}\,\Gamma(\nu)}.
\end{equation}
In practice, equation~\eqref{eq:SPDE} is solved on a bounded domain with arbitrary boundary conditions~\cite{daon2018mitigating,graham2018analysis,khristenko2018analysis}.

In the general case, the shape operator can be spatially varying: $\tns{\Theta}= \tns{\Theta}(\x)$ (see, for example, \cite{roininen2014whittle}),
which leads to a non-stationary covariance.
As mentioned above, we consider in this work the stationary covariance case and thus the spatially invariant shape operator~$\tns{\Theta}$.
This allows the solution of the equation~\eqref{eq:SPDE} using the Fourier Transform:
\begin{equation}\label{eq:solution_by_Fourier}
	\Int(\x; \omega) = \Fi\left\lbrace \hat{G}(\vct{\xi})\cdot\WNhat(\vct{\xi};\omega) \right\rbrace(\x),
\end{equation}
where $\Fi$ denotes the operator of the inverse Fourier Transform, $\WNhat$ is the Fourier Transform of the white noise~$\WN$, and $\hat{G}$ is the Fourier Transform of the Green's operator for the operator in~\eqref{eq:SPDE}, which is given by (see \cite[Vol.II, section~$8.13$, formula~$(3)$]{bateman1954tables})
\begin{equation}\label{eq:MaternFourier}
	\abs{\hat{G}(\vct{\xi})}^2 = \frac{(2\pi)^{\frac{d}{2}}\sqrt{\det\tns{\Theta}}\,\Gamma(\nu+d/2)}{\nu^{\frac{d}{2}}\,\Gamma(\nu)}
	\cdot \left(1 + \frac{1}{2\nu}\,\vct{\xi}\cdot {\tns{\Theta}}\vct{\xi} \right)^{-(\nu + \frac{d}{2})}.
\end{equation}
Use of the Fast Fourier Transform algorithm in convolution~(\ref{eq:solution_by_Fourier}), along with periodic boundaries, makes possible a fast generation of the synthetic samples.



\begin{figure}[!ht]
	\centering
	\newcommand{\size}{0.3\textwidth}
	
	\begin{subfigure}{\size}
		\centering
		\includegraphics[height=\textwidth]{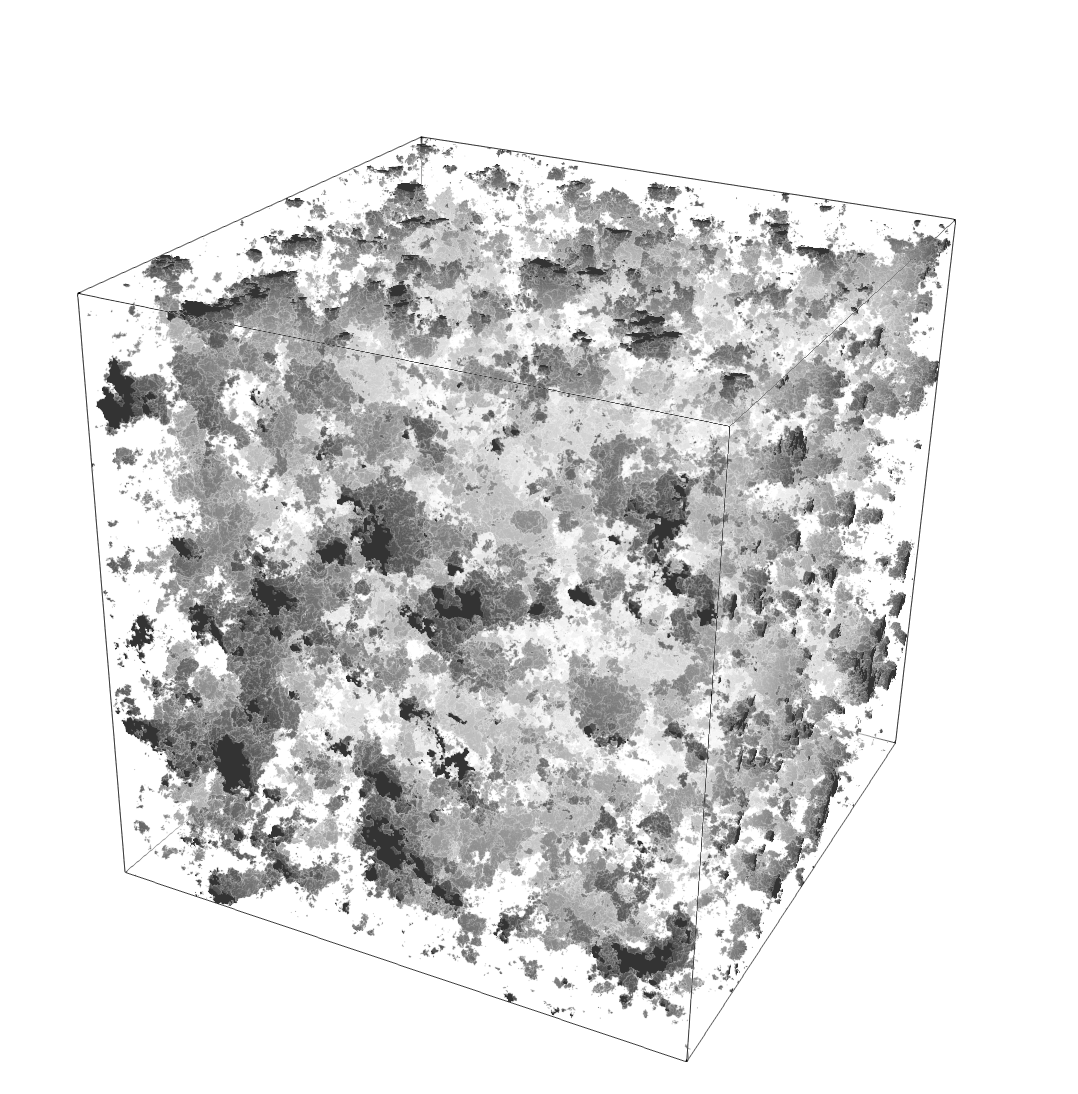}
		\caption{\scriptsize $\vfm = 0.05$, $\nu = 0.5$, \\ $\corrlen_i=0.05$.}\label{sbfig:example1}
	\end{subfigure}
	\hfill
	\begin{subfigure}{\size}
		\centering
		\includegraphics[height=\textwidth]{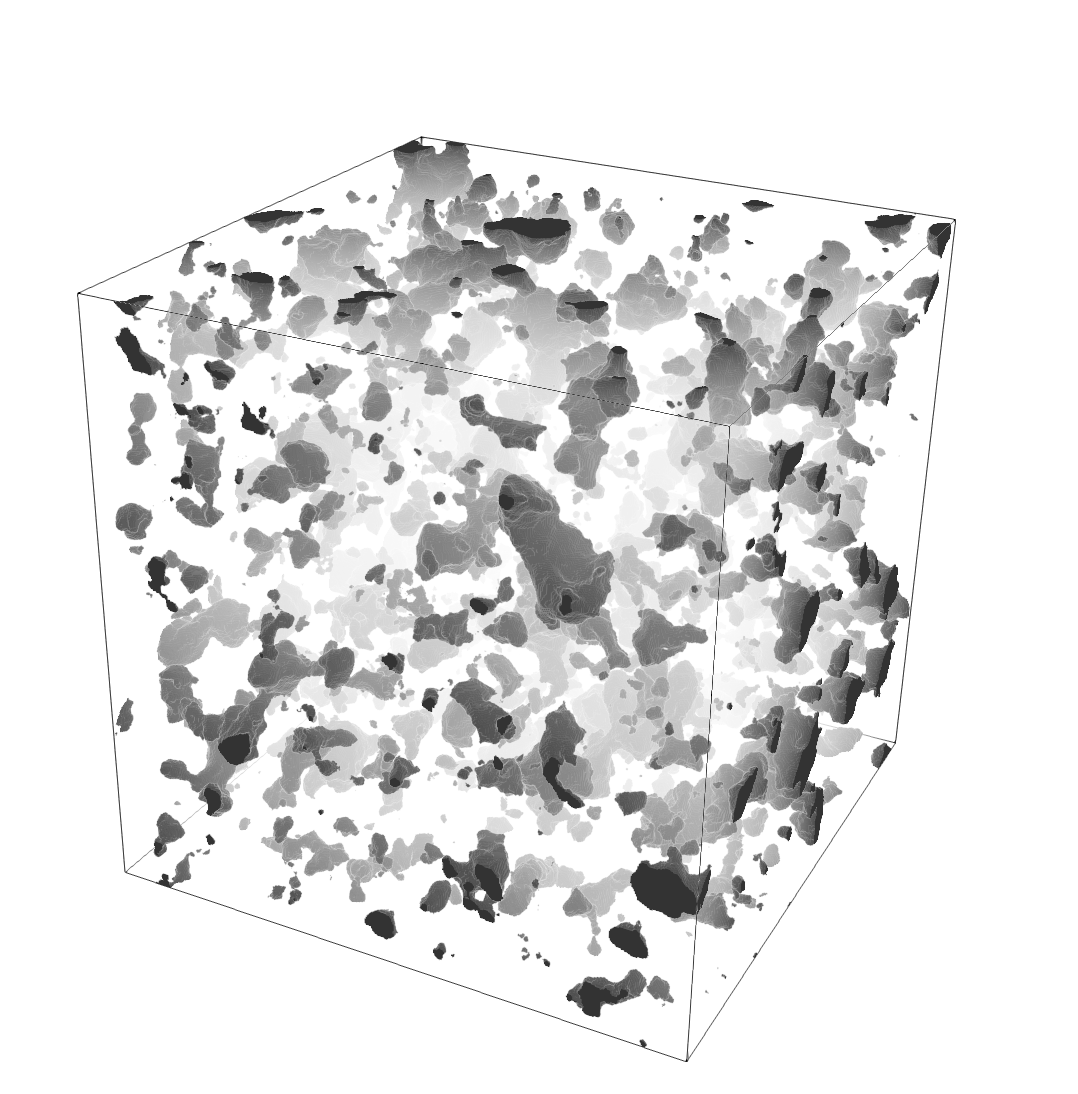}
		\caption{\scriptsize $\vfm = 0.05$, $\nu = 1.5$, \\ $\corrlen_i=0.05$.}\label{sbfig:example2}
	\end{subfigure}
	\hfill
	\begin{subfigure}{\size}
		\centering
		\includegraphics[height=\textwidth]{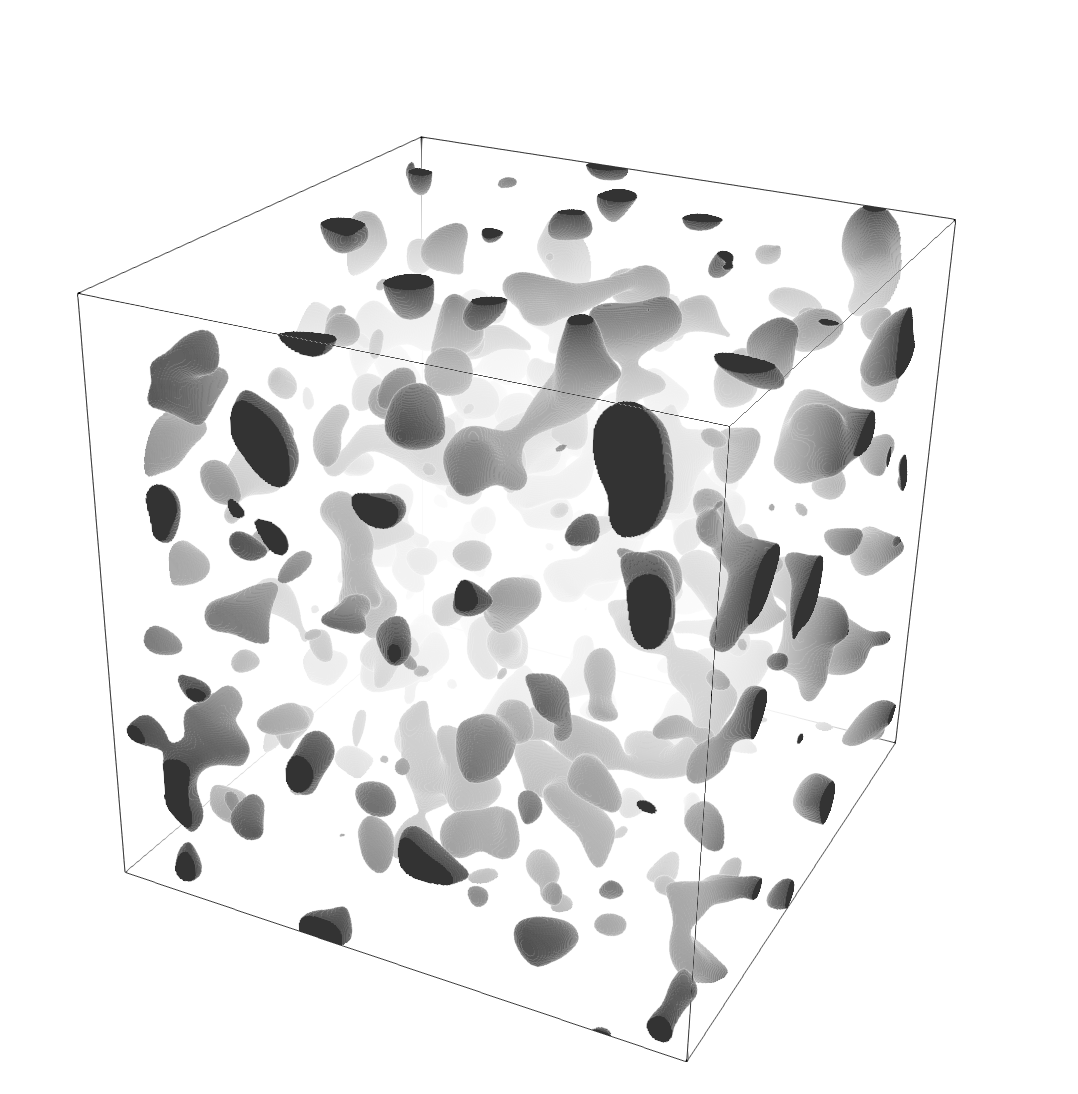}
		\caption{\scriptsize $\vfm = 0.05$, $\nu = 10$, \\ $\corrlen_i=0.05$.}\label{sbfig:example3}
	\end{subfigure}
	
	\vspace{1ex}
	
	\begin{subfigure}{\size}
		\centering
		\includegraphics[height=\textwidth]{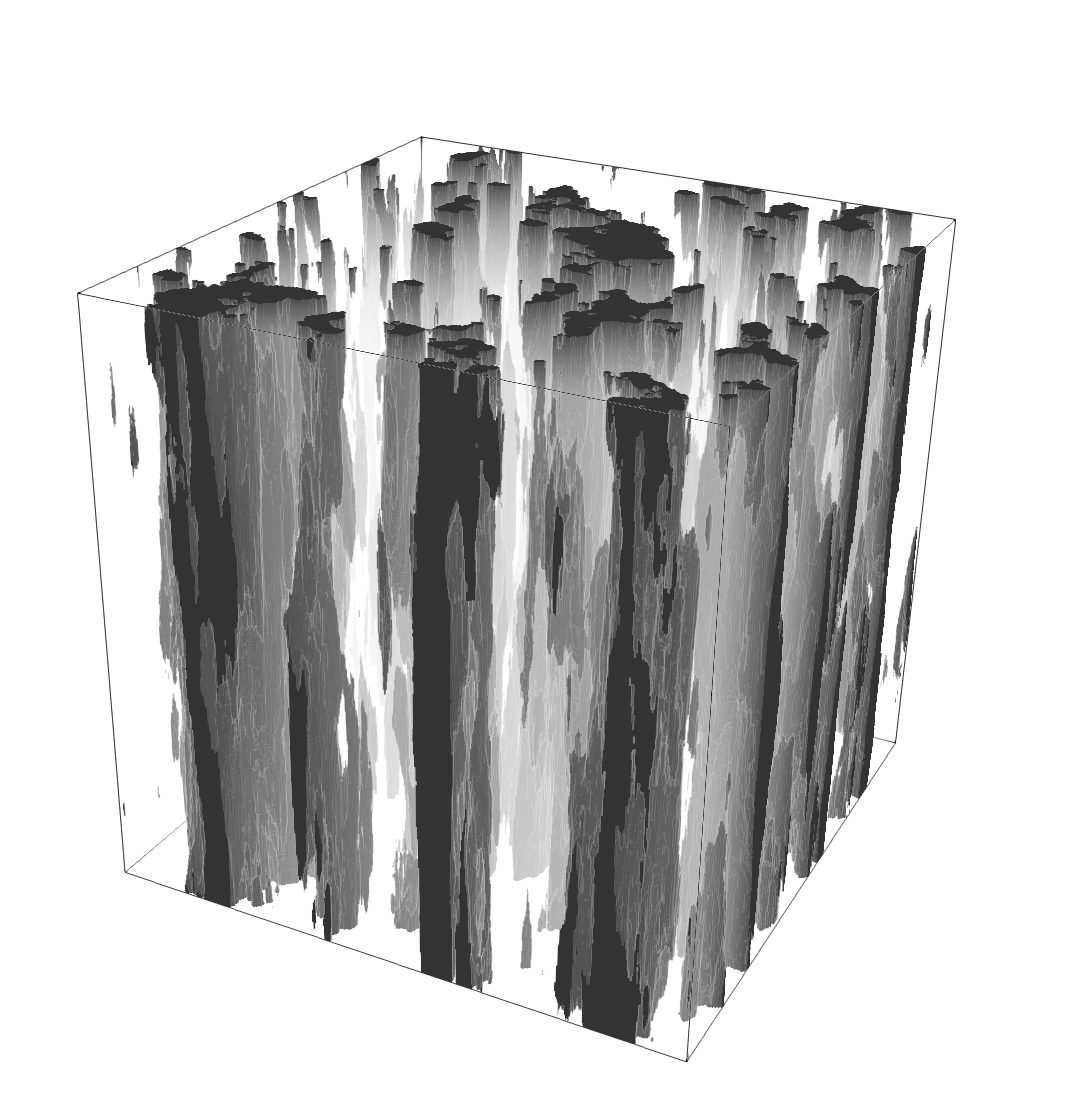}
		\caption{\scriptsize $\vfm = 0.1$, $\nu = 1$, \\ $0.1\cdot\corrlen_1=\corrlen_2=\corrlen_3=0.05$.}\label{sbfig:example5}
	\end{subfigure}
	\hfill
	\begin{subfigure}{\size}
		\centering
		\includegraphics[height=\textwidth]{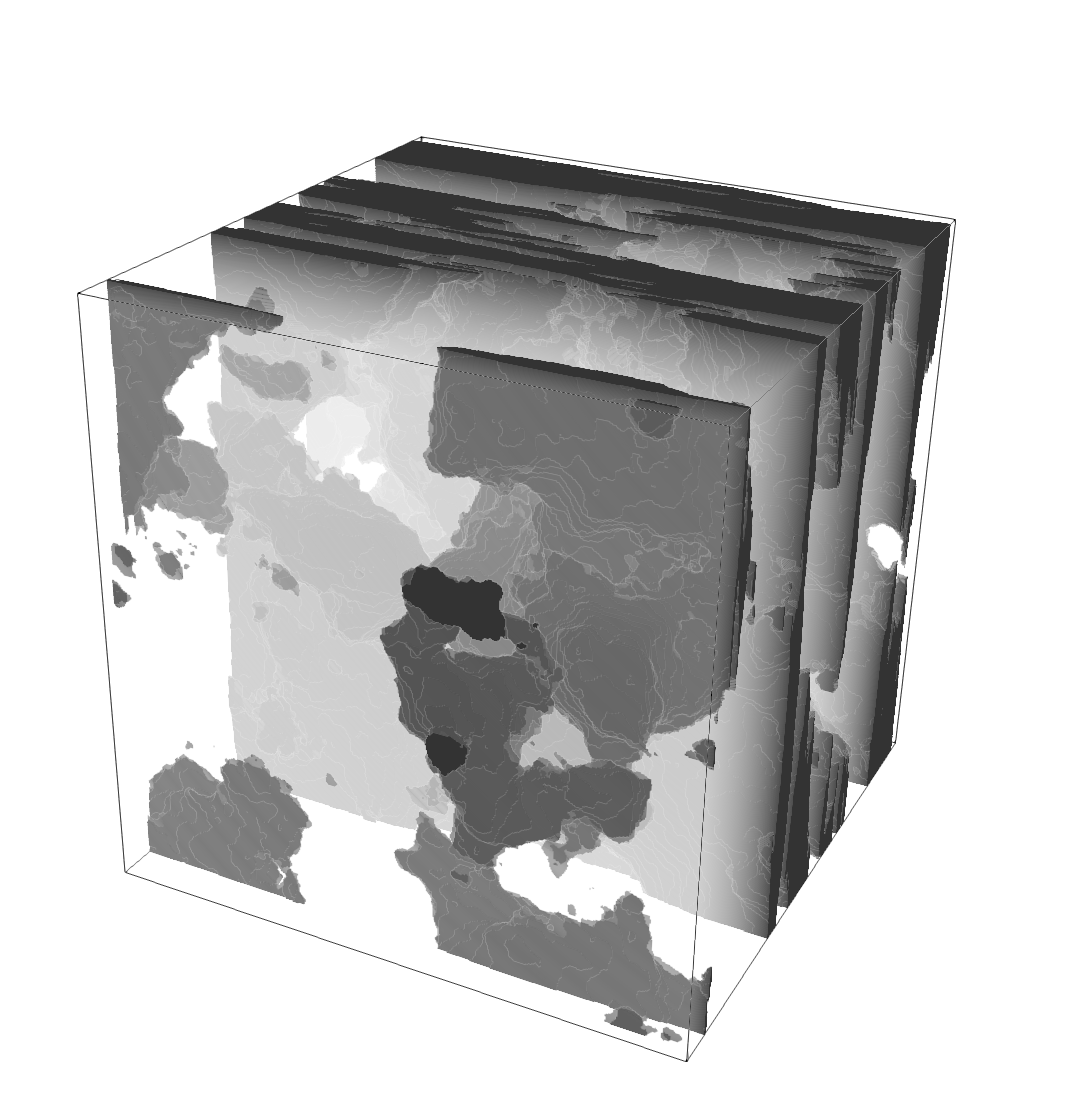}
		\caption{\scriptsize $\vfm = 0.1$, $\nu = 1$, \\ $0.1\cdot\corrlen_1=0.1\cdot\corrlen_2=\corrlen_3=0.05$.}\label{sbfig:example6}
	\end{subfigure}
	\hfill
	\begin{subfigure}{\size}
		\centering
		\includegraphics[height=\textwidth]{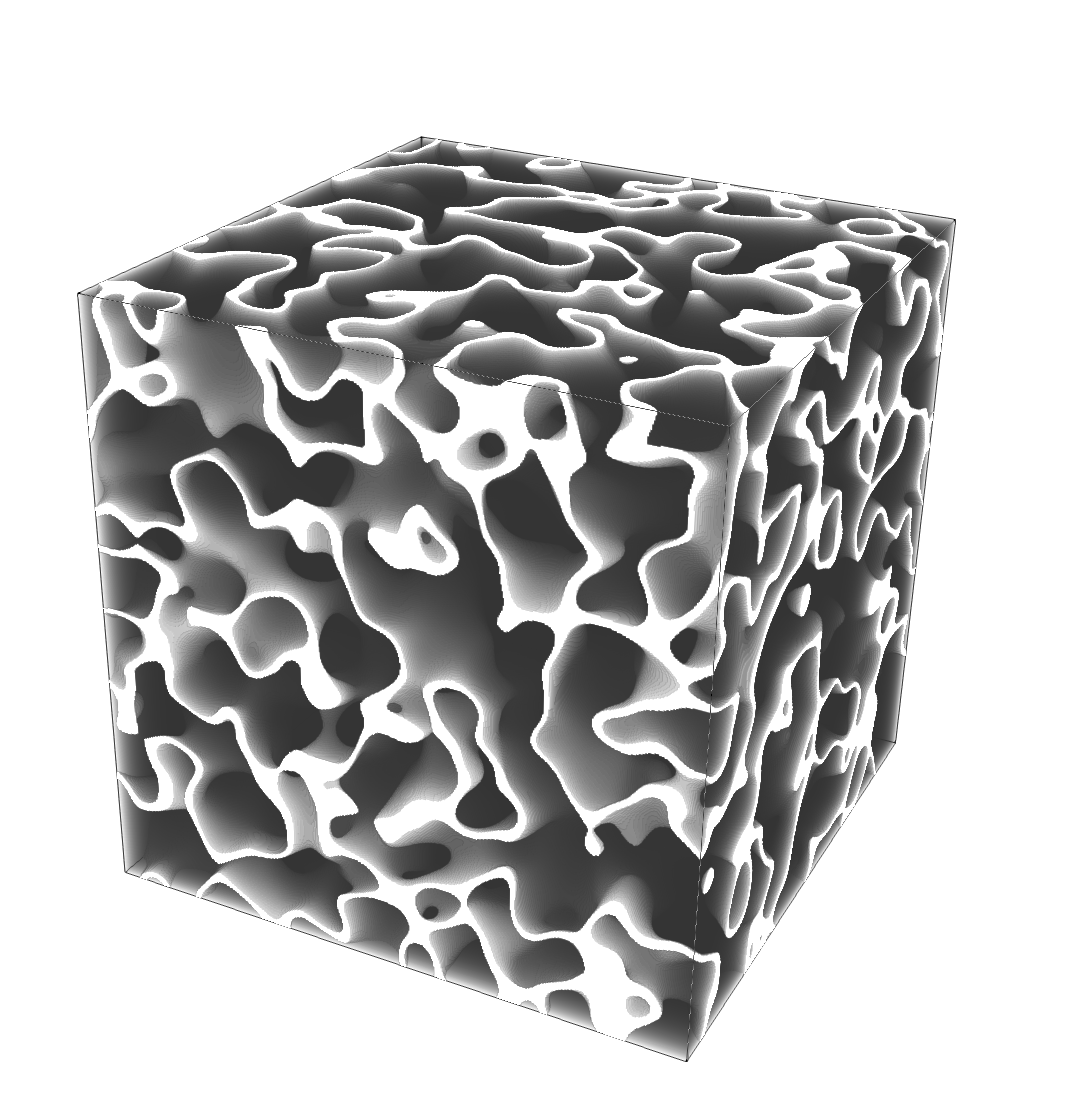}
		\caption{\scriptsize $\vfm = 0.8$, $\nu = 10$, \\ $\corrlen_i=0.05$.}\label{sbfig:example4}
	\end{subfigure}
	
	\caption{Examples of surrogate material. (\subref{sbfig:example1})-(\subref{sbfig:example3}) influence of the parameter~$\nu$: with porosity~$\vfm$ and correlation lengths~$\corrlen_i$ fixed, the smoothness of the pores increases with~$\nu$. (\subref{sbfig:example5})-(\subref{sbfig:example6}) anisotropic media examples. (\subref{sbfig:example4}) high porosity value example.
	}		
	\label{fig:examples}
\end{figure}

\vspace{2ex}
Thus, given design parameters~$\vfm$ (or $\tau$), $\nu$ and $\tns{\Theta}$, the process of generating a microstructure realization for the surrogate material model described above can be summarized as follows:
\begin{enumerate}
	\item Draw a realization of the spatial white noise~$\WN(\x)$.
	\item Compute the corresponding intensity field~$\Int(\x)$ by solving~\eqref{eq:SPDE}.
	For the stationary case, it is given by convolution~\eqref{eq:solution_by_Fourier}, computed using FFT.
	\item The characteristic function~$\Phase(\x)$ of the inclusions phase  is given by the level-cut~\eqref{eq:levelcut} of the intensity~$\Int(\x)$.
\end{enumerate}
This surrogate material model is completely determined by only a few design parameters~$\vfm$, $\nu$ and $\tns{\Theta}$.
However, by varying these parameters, the model can produce a large number of classes of random heterogeneous media~(see examples in Figure~\ref{fig:examples}).
Besides, the parameters have clear intuitive meanings:
\begin{itemize}
	\item The parameter~$\vfm$ is the expected volume fraction of the inclusions (or porosity).
	\item The regularity parameter~$\nu$ defines the differentiability (smoothness) of the inclusions interface.
	\item The shape operator~$\tns{\Theta}$ defines the metric to measure the distance in the covariance function and thus controls the form of the inclusions: their size, anisotropy level (aspect ratio) and orientation.
\end{itemize}

\begin{rmk}
	In~\cite{bolin2018numerical}, it is shown that, given the white noise is almost sure in~$\H^{-d/2-\epsilon}(\R^d)$, $\epsilon>0$, the solution of~(\ref{eq:SPDE}) is in $\H^{\nu-\epsilon}(\R^d)$.
	Then, from the Sobolev embedding theorem, the intensity field~$\Int(\x)$ is continuous for $\nu>d/2$.
	And for $\nu\le d/2$, we deal with a discontinuous intensity field~$\Int(\x)$, which produces a disjoint particles microstructure (see Figure~\ref{sbfig:example1}).
\end{rmk}

\begin{figure}[!ht]
	\centering
	
	\newcommand{\markscale}{0.5}
	
	\begin{tikzpicture}[scale=1.0]
	
		\begin{groupplot}[
			axis standard,
			width=0.49\textwidth,
			group style={group name=myplot,group size= 2 by 1, horizontal sep=2cm,},
			cycle list/Paired,			
		]
		
			\nextgroupplot[
				ylabel={$\RelSize$},
				xlabel={Sphericity},
				legend columns=5,
				legend style={at={(1.25,-0.4)},anchor=north,font=\scriptsize},
				title={$\vfm=0.05$},
				cycle list shift=1
			]

				\pgfplotstableread[header=true, col sep=comma]{data_files/plot_PoreGeometry_vf=0.05.csv}{\data}	
				\pgfplotstablegetcolsof{\data}\pgfmathsetmacro{\Nnu}{int(\pgfplotsretval/2)}
				
				
				\pgfplotsforeachungrouped \i in {1,...,\Nnu} {
					\pgfmathtruncatemacro{\Rindex}{int(2*\i-2)}
					\pgfmathtruncatemacro{\Sindex}{int(2*\i-1)}
					\pgfmathtruncatemacro{\redfrac}{(\i-1)*100/(\Nnu-1)}
					\edef\temp{
						\noexpand\addplot+[mark=o, only marks, mark options={scale=\markscale}, thick] table[x index=\Sindex, y index=\Rindex, header=true, col sep=comma] {\noexpand\data};
					}\temp
					\pgfplotstablegetcolumnnamebyindex{\Rindex}\of{\data}\to\nuval
					\addlegendentryexpanded{ $\nu=\nuval\quad$ }
				}
		
			\nextgroupplot[
				ylabel={$\RelSize$},
				xlabel={Sphericity},
				title={$\vfm=0.2$},
				cycle list shift=1
			]		
			
				\pgfplotstableread[header=true, col sep=comma]{data_files/plot_PoreGeometry_vf=0.20.csv}{\data}	
				\pgfplotstablegetcolsof{\data}\pgfmathsetmacro{\Nnu}{int(\pgfplotsretval/2)}
				
				
				\pgfplotsforeachungrouped \i in {1,...,\Nnu} {
					\pgfmathtruncatemacro{\Rindex}{int(2*\i-2)}
					\pgfmathtruncatemacro{\Sindex}{int(2*\i-1)}
					\pgfmathtruncatemacro{\redfrac}{(\i-1)*100/(\Nnu-1)}
					\edef\temp{
						\noexpand\addplot+[mark=o, only marks, mark options={scale=\markscale}, thick] table[x index=\Sindex, y index=\Rindex, header=true, col sep=comma] {\noexpand\data};
					}\temp
				}
		\end{groupplot}
	\end{tikzpicture}
	
	\caption{Sphericity as a function of the relative pore size~$\RelSize$, eq. (\ref{eq:Sphericity}), observed on $500$ pores, with porosity~$\vfm=0.05$~(left) and $\vfm=0.2$~(right) and varying~$\nu$ (marked by color).
	Computed for 3D surrogate samples of about $10^8$ voxels.}
	\label{fig:PoreGeometry}
\end{figure}
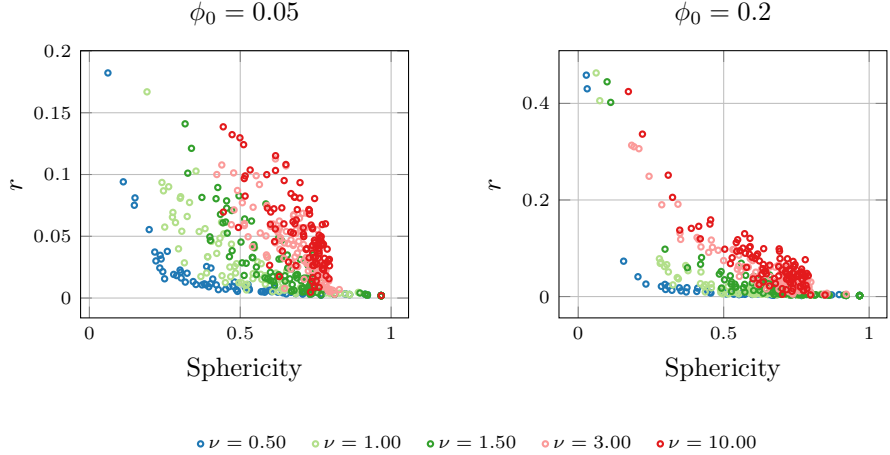

The regularity parameter~$\nu$ is also related to the pore sphericity, one of the important parameters describing a microstructure~\cite{buffiere2001experimental,le2017,zhou2017}.
We use the sphericity definition, proposed in~\cite{wadell1935volume}, and define the relative pore size~$\RelSize$:
\begin{equation}\label{eq:Sphericity}
	\Sphericity = \frac{\pi^{1/3} (6\,\Vpore)^{2/3}}{0.833\cdot\Spore}, \qquad
	\RelSize = \left(\frac{\Vpore}{\Vtot}\right)^{\frac{1}{3}},
\end{equation}
where $\Vpore$ and $\Spore$ denote the volume of the pore and the surface of the pore interface respectively, and $\Vtot$ is the total volume of the sample.
A correction factor of~$0.833$ is taken into account due to the digitized structure of the surface~\cite{zhou2017}.
Figure~\ref{fig:PoreGeometry} shows the distribution of the relative pore size and the pore sphericity, computed for 3D surrogate samples of $2^{9\cdot 3}\approx 10^8$ voxels.
We observe that with growing size the sphericity depends on~$\nu$: the larger~$\nu$ is, the more spherical are the pores.
With the correlation length fixed, higher porosity~$\vfm$ provides larger maximum pore size.

We use the Mat\'ern covariance class because of its flexibility, i.e. it covers a large variety of morphologies (microstructural shapes) with a relatively simple parametrization.
Nevertheless, other convenient covariance kernel can be used instead (see, for example,~\cite{matern1986spatial,stein2012interpolation,lim2009generalized,teubner1991level}).
In the general case, in the absence of an explicit formula for the power spectrum, like~(\ref{eq:MaternFourier}), it can always be directly computed using FFT.



\subsection{Bayesian inference of the model parameters}
\label{sec:Reconstruction}

Let us suppose that several binary images (realizations~$\omega$) $\Phase(\x; \omega)$ of a real material are given.
Under the stationarity assumption, we can approximate the two-point probability function~$S_2(r)$ of this material with the spatial average
\begin{align}\label{eq:S2_data}
S_2^{data}(\vct{r}_n) = \frac{1}{N_n}\sum_{k}\Phase(\x_k)\,\Phase(\x_k+\vct{r}_n),
\end{align}
where $\x_k$ runs through all the voxels in all the provided images, $\Phase(\x)$ being zero outside the images (zero padding), and $N_n$ is the number of points $\x_k+\vct{r}_n$ belonging to the images.
We compute the cross-correlation~(\ref{eq:S2_data}) using FFT with zero padding.

The two-point probability function $S_2(r)=\E{\Phase(\x;\omega)\Phase(\x+\vct{r};\omega)}$, $r=\norm{\vct{r}}$, is the probability of the event when two points separated with distance~$r$ are in the same phase.
Then, the random variable $z_k(\vct{r};\omega) = \Phase(\x_k;\omega)\Phase(\x_k+\vct{r};\omega)$ has the Bernoulli distribution with parameter $p=S_2(r)$.
In~(\ref{eq:S2_data}) we compute the sum of "trials" of $z(r;\omega,\x)$, which is, thus, from the Binomial distribution.
Since this tends with a large number of trials to the normal distribution~$\Gaussian\left(p, \frac{p(1-p)}{N}\right)$ with the probability density function, we have
\begin{equation}
	f\left(\frac{1}{N}\sum_{k=1}^{N}z_k\right) \approx \frac{1}{\sqrt{2\pi\,N^{-1}\,p\cdot(1-p)}}\Exp{-\frac{\left(\frac{1}{N}\sum_{k=1}^{N}z_k - p\right)^2}{2\,N^{-1}\,p\cdot(1-p)}},
	\qquad \forall\vct{r}_n.
\end{equation}
Thus, the product of the densities for all $\vct{r}_n$ provides the likelihood for our Bayesian inverse problem:
\begin{equation}\label{eq:Likelihood}
	\Likelihood = \prod_{n}\frac{\operatorname{Constant}_n}{\sqrt{S_2(r_n)\cdot(1-S_2(r_n))}}\Exp{-\frac{\left(S_2^{data}(\vct{r}_n) - S_2(r_n)\right)^2}{2\,N_n^{-1}\,S_2(r_n)\cdot(1-S_2(r_n))}},
\end{equation}
where we can substitute the real two-point correlation~$S_2(r)$ within our surrogate model.
The surrogate two-point probability function is given by formula~(\ref{eq:S2}):
\begin{equation}
S_2^{model}(r; \prms) = 2\vfm - 4T\left(\tau, \sqrt{\frac{1-\cov(r;\prms)}{1+\cov(r;\prms)}} \right) - 4T\left(\tau, \sqrt{\frac{1+\cov(r;\prms)}{1-\cov(r;\prms)}} \right)
\end{equation}
where $\cov(r;\prms)$ is the intensity field covariance kernel, $\vfm$ is the average volume fraction of the pores (porosity), $\tau$ is given by~(\ref{eq:tau_vf}), and $\prms$ denotes the vector of design parameters of the model.
In our case, $\cov(r;\prms)$ is of Mat\'ern type in the form~(\ref{eq:MaternCovariance}) with unit variance~$\sigma=1$, and $\prms=(\vfm, \nu, \Shape)$.

Assuming a uniform prior, the posterior distribution is simply determined by the likelihood.
We use the Laplace approximation of the posterior: we approximate it with a multivariate lognormal distribution~$\log\Gaussian(\log\prms_{MLE},\tns{\Sigma})$, where
the maximal likelihood estimator~$\prms_{MLE}$ is given by
\begin{equation}\label{eq:reconst:argmax}
	\prms_{MLE} = \arg\max_{\prms\in\prmspace} \Likelihood(\prms),
\end{equation}
and the covariance matrix is approximated with
\begin{equation}\label{eq:reconst:hess}
	\tns{\Sigma} = -\left.\left(\frac{\d^2 \log\Likelihood(\prms)}{\d\,(\log\prms)^2}\right)\inv\right|_{\prms=\prms_{MLE}}.
\end{equation}

In the isotropic case, we have $\Shape=\corrlen^2\,\tns{\Id}$.
Then $\corrlen$ simply plays the role of the characteristic length scale and does not affect the microstructure morphology.
So, we simply fix $\corrlen\equiv\corrlen_{MLE}$ as a deterministic parameter.

	\section{Fatigue Analysis: mechanical formulation}\label{sec:LE}

In what follows, we will perform a mechanical analysis on the surrogate microstructure with periodic boundary conditions under a given mean cyclic loading.
The output of the computation will be analyzed in terms of quantities of interest for a standard high cycle fatigue analysis.
The goal is to estimate the uncertainty of these quantities with respect to the variability of the underlying surrogate microstructures.
From the point of view of a fatigue design method, the surrogate microstructure represents the material RVE and carries the statistical features of the real material as discussed in the previous sections.
Moreover, the RVE represents a material point of a large structure and the given cyclic loading is the local shakedown response of the structure under service loading. 
In order to simplify the discussion and the subsequent computations, we will next only consider an elastic material behavior for the RVE.
This assumption does however not restrain the generality of the present method and can be applied to materials with plasticity, viscosity, damage, etc., and can equally cover multiples material phases.
  
\subsection{Linear elasticity problem}
\label{sec:Problem}

Let $D=(0,1)^3$ be a representative volume element (RVE) of a heterogeneous two-phase material.
The phases are defined by a stochastic characteristic function~$\Phase(\x; \omega)$ as in~(\ref{eq:char}),  periodic on~$D$.
The material is submitted to the macroscopic stress~$\Macrostress$ to be imposed on average on the RVE, which results in a macroscopic strain~$\Macrostrain$.
Thus, for all $\x\in\R^d$, the total displacement field~$\disp(\x)$ is of the form~\cite{suquet1985elements}
\begin{equation}\label{eq:PeriodicBC}
	\disp(\x) = \Macrostrain\cdot\x + \tilde{\disp}(\x),
\end{equation}
where the fluctuation~$\tilde{\disp}(\x)$ is periodic on~$D$.

Under the assumption of small strains, the local microscopic strain tensor field, observed on the stochastic RVE after loading, is
\begin{equation}
	\strain(\x) = \frac{1}{2}\left(\grad\disp(\x) + \grad\disp(\x)\tp\right),
\end{equation}
and then condition~(\ref{eq:PeriodicBC}) can be written in terms of strains:
\begin{equation}\label{eq:PeriodicBC_strain}
	\strain(\x) = \Macrostrain + \tilde{\strain}(\x), \qquad \tilde{\strain}(\x) = \frac{1}{2}\left(\grad\tilde{\disp}(\x)+\grad\tilde{\disp}(\x)\tp\right),
\end{equation}
where $\tilde{\strain}(\x)$ has zero mean and is periodic on $D$, and $\Macrostrain$ denotes the macrostrain.

The local stress tensor field~$\stress(\x)$ is
\begin{equation}
\stress(\x) = \Stiffness(\x)\cdot\strain(\x) = \bbc{\bulk(\x)-\tfrac{2}{d}\shear(\x)}\,\tr(\strain(\x))\;\tns{Id} + 2\shear(\x)\,\strain(\x),
\end{equation}
where $\Stiffness(\x)$ is the fourth order stiffness tensor, $\bulk(\x)$ and $\shear(\x)$ are the bulk and shear moduli, respectively, which are defined by the phase~$\Phase(\x)$:
\begin{equation}
\bulk(\x) = \bulk_I\,\Phase(\x) + \bulk_M\,(1-\Phase(\x)), \qquad \shear(\x) = \shear_I\,\Phase(\x) + \shear_M\,(1-\Phase(\x)),
\end{equation}
where $\bulk_I, \shear_I$ are the bulk and shear moduli of the inclusions, and $\bulk_M, \shear_M$ -- of the matrix, respectively.
The equilibrium is given by
\begin{gather}\label{eq:LE system}
	\div\stress(\x) = 0 \quad\text{in }D, \qquad \int\limits_{D}\stress(\x) \d\x = \Macrostress,
\end{gather}
where $\Macrostress$ denotes the macrostress.

Thus, for a given macrostress~$\Macrostress$, we want to find the local strain field $\strain(\x)=\Macrostrain + \tilde{\strain}(\x)$, such that the macrostrain~$\Macrostrain$ and the fluctuation $\tilde{\strain}(\x)=\frac{1}{2}\left(\grad\tilde{\disp}(\x)+\grad\tilde{\disp}(\x)\tp\right)$ are the solution of the system
\begin{gather}\label{eq:LE in strains}
	\div\bigl(\Stiffness(\x)\cdot\tilde{\strain}(\x)\bigr) = -\div\bigl(\Stiffness(\x)\cdot\Macrostrain\bigr) \quad\text{in }D, \\
	\int\limits_{D}\Stiffness(\x)\cdot(\Macrostrain+\tilde{\strain}(\x)) \d\x = \Macrostress, \\
	\tilde{\strain}(\x) \text{ is periodic on D}. 
\end{gather}

\subsection{Mechanical Quantity of Interest for multidimensional loading}
\label{sec:QoI}

The high cycle fatigue analysis of metals is based on an elastic shakedown assumption at the structural scale and on material observations at the microscopic scale. 
Moreover, plasticity is driven by the deviatoric part of stresses, and cracks and porosities are opened by the positive spherical part of the stresses, which provides a natural split in the stress space. 
Therefore, a characteristic loading path can be characterized by a combination of bulk and shear loading as follows:
\begin{align}\label{eq:loading}
	\Macrostress_\theta
	&= \left(\begin{array}{ccc} \frac{1}{\sqrt{3}} \cos \theta  & - \frac{1}{\sqrt{2}} \sin \theta & 0 \\ -\frac{1}{\sqrt{2}} \sin \theta & \frac{1}{\sqrt{3}} \cos \theta & 0 \\0&0 & \frac{1}{\sqrt{3}} \cos \theta \end{array}\right) \\
	&= \cos\theta \cdot \left(\begin{array}{ccc} \frac{1}{\sqrt{3}} & 0 & 0 \\ 0 & \frac{1}{\sqrt{3}} & 0 \\0&0 & \frac{1}{\sqrt{3}} \end{array}\right) 
	+ \sin\theta \cdot \left(\begin{array}{ccc} 0 & - \frac{1}{\sqrt{2}} & 0 \\ - \frac{1}{\sqrt{2}} & 0 & 0 \\0&0 & 0 \end{array}\right),
\end{align}
where the angle $\theta$ varies from $0$ to $\pi/2$ and characterizes the share of hydrostatic and deviatoric loading respectively.
Lode coordinates of this loading are $z=\cos \theta$, $r=\sin \theta$ and $\theta_s =0$.
In the stress space, the loading is a unit vector in the meridional plane $(z,r)$ and  varies continuously from an isotropic traction  with eigenstress  $(\frac{1}{\sqrt{3}},\frac{1}{\sqrt{3}},\frac{1}{\sqrt{3}})$ obtained at $\theta=0$ to a pure shear with eigenstress $(-\frac{1}{\sqrt{2}},0,\frac{1}{\sqrt{2}})$ obtained at $\theta = \pi/2$.

We can compute for a given microstructure~$\Phase(\x; \omega)$, $\x\in D$, the elastic solutions 
$\stress_0(\x; \omega)$ and  $\stress_{\pi/2}(\x; \omega)$ corresponding to the loads $\Macrostress_0$ and 
$\Macrostress_{\pi/2}$, respectively.
We decompose the local stress field~$\stress(\x)$ into a deviatoric part~$\tns{\dv} = \stress-\frac{1}{3}\tr(\stress)\tns{\Id}$ and a spherical part~$\frac{1}{3}\tr(\stress)\tns{\Id}$, and introduce
the local "traction"~$\tr_+(\stress) = \max(\tr(\stress), 0)$.
Then, for each angle $\theta$, we can obtain the associated quantities
\begin{equation}
	\tr_+(\stress_\theta(\x;\omega))= \tr_+\biggl(\cos \theta \cdot \stress_0 + \sin \theta \cdot \stress_{\pi/2} \biggr)(\x; \omega)
\end{equation}
and
\begin{equation}
	\dv_\theta(\x;\omega)= \biggl(\cos \theta \cdot \dv_0 + \sin \theta \cdot  \dv_{\pi/2}\biggr)(\x; \omega)
\end{equation}
and deduce the associated combined quantity of interest over the volume~$D$
\begin{equation}\label{eq:QoI}
	\QoI(\omega,\theta) = \sqrt{\frac{1}{|\nbhd_t(\omega,\theta)|}\int\limits_{\nbhd_t(\omega,\theta)} d(\x;\omega,\theta)^2 d\x},
\end{equation}
where
\begin{equation}\label{eq:damage}
	d(\x;\omega,\theta) = \norm{\dv_\theta(\x;\omega)} + 0.3\cdot\tr_+(\stress_\theta(\x;\omega))
\end{equation}
is the microscopic damage parameter, and $\norm{\dv}=\sqrt{\dv:\dv}$.
This weighted combination of deviatoric norm and of positive trace  is often used in practical fatigue criteria~\cite{hofmann2009numerical}.
The slope~$0.3$ is an empirical value.
The subdomain of integration $O_t(\omega,\theta)$ is here a ball of radius $\corrlen$ centered at the point~$\x$ which maximizes~$d(\x;\omega,\theta)$, the pores being excluded.

We are interested in the expected value $\E{\QoI(\theta)}$ and the variance $\Var{\QoI(\theta)}$, which are functions of~$\theta$, and the "worst" loading  combination will be the one leading to the maximum expected value~$\max\limits_{0\le\theta\le\frac{\pi}{2}}\E{\QoI(\theta)}$.

The quantity~$\QoI$ presents the mean squared damage parameter around the maximum point.
Since, in general, the maximum point can be a singularity, the integral~\eqref{eq:QoI} of the damage parameter is preferred as quantity of interest.
Having the estimation of the damage parameter in hand, one can proceed with the analysis of the number of loading cycles to failure~\cite{buffiere2001experimental,hofmann2009numerical} or the failure probability~\cite{charkaluk2014probability}.

	\section{Fatigue Analysis: results and discussions}\label{sec:Sim}

In this section, we apply the surrogate microstructure model, discussed in Section~\ref{sec:Model}, to the linear elasticity problem, formulated in Section~\ref{sec:LE}, in order to perform a risk analysis of a heterogeneous material with respect to fatigue.
Access to the fast microstructure sampling algorithm allows the approximation of the probability distribution of the fatigue criteria using Monte-Carlo methods.
Furthermore, we can construct in this way a mapping from the microstructure design parameters to quantities of interest, which can serve for analysis of the influence of the microstructure morphology properties on the quantities of interest.
Thus, in this section, we discuss two types of results:
\begin{itemize}
	\item First, using the Laplace approximation for Bayesian inference of the microstructure design parameters (see Section~\ref{sec:Reconstruction}), a surrogate material is reconstructed from a few CT images of a real material.
	The statistical properties of the simplified fatigue criteria~(\ref{eq:QoI}) for the surrogate material are analyzed using the Monte-Carlo method, employing a fast Fourier-based sampling process and linear elasticity solver.	
	
	\item Second, we perform the sensitivity analysis of the fatigue criteria and the homogenized elastic moduli with respect to the design parameters: porosity~$\vfm$ and pore regularity~$\nu$.
	This makes possible the optimization of the microstructure design with respect to fatigue.
\end{itemize}

All simulations have been performed using an Intel Xeon E5 processor ($48$x~$3.00$GHz) and $504$ GB RAM, computing Monte-Carlo samples in parallel.
Each Monte-Carlo iteration includes:
\begin{enumerate}
	\item Sample surrogate microstructure using Mat\'ern covariance kernel (Section~\ref{sec:Model}).
	\item Solution of the linear elasticity problem~(\ref{eq:LE in strains}) for the loadings~$\theta=0$ and $\theta=\pi/2$ to obtain the local strains and stresses for the current microstructure sample (Section~\ref{sec:LE}).
	Given the problem linearity, the results for intermediate loads~$\theta\in(0,\pi/2)$ are computed as linear combination of the cases $\theta=0$ and $\theta=\pi/2$.
	\item Computation of quantities of interest for the current sample (Subsection~\ref{sec:QoI}).
\end{enumerate}
Runtime of one such iteration is about~$1.5\,s$ for a 2D sample of $2^{8\cdot2}\approx 6.5\cdot 10^4$ voxels, or about~$70\,s$ for a 3D sample of $2^{7\cdot 3}\approx {2\cdot 10^{6}}$ voxels.

The linear elasticity problem~(\ref{eq:LE in strains}) is solved in terms of local strains, using a Fourier-based Krylov solver, proposed by~\cite{brisard2010fft}, which is a modification of the original method of~\cite{moulinec1994fast}, providing the convergence in case of the infinite contrast.
See also~\cite{moulinec1998numerical,michel2001computational}.
The problem is implemented with an in-house code written in Python and C++, using FFTW~\cite{FFTW05} and PETSc~\cite{petsc-web-page} libraries for the Fast Fourier Transform and the matrix-free conjugated gradient method respectively.
{\it SciPy}~\cite{scipy} and {\it Numdifftools}~\cite{numdifftools} packages have been used to solve the optimization problem~\eqref{eq:reconst:argmax} and to compute the hessian~\eqref{eq:reconst:hess} respectively.



\subsection{Fatigue analysis of a real material}
\label{sec:AppRealMat}

\begin{figure}[!hp]
	\newcommand{\sizemat}{0.2\textwidth}
	\newcommand{\sizesgt}{0.2\textwidth}
	\newcommand{\sizesgtv}{0.4\textwidth}
		
	\begin{subfigure}{\textwidth}
		\centering
		\makebox[\textwidth]{
			\frame{\includegraphics[height=\sizemat]{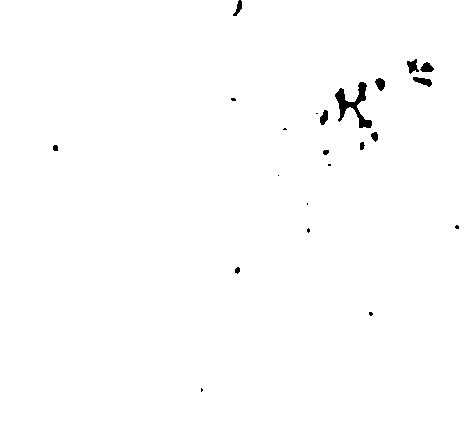}}
			\frame{\includegraphics[height=\sizemat]{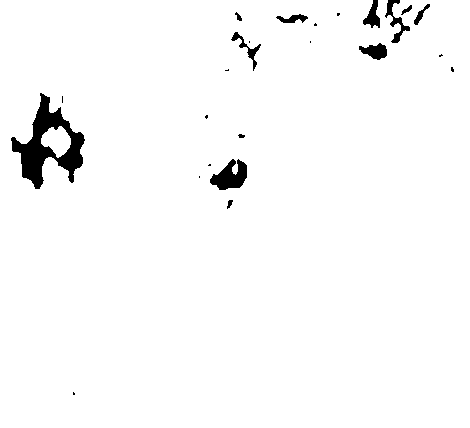}}
			\frame{\includegraphics[height=\sizemat]{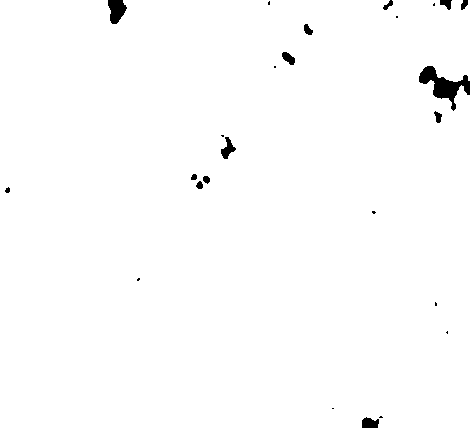}}
		}
	
		\vspace{0.5ex}	
		\makebox[\textwidth]{
			\frame{\includegraphics[height=\sizemat]{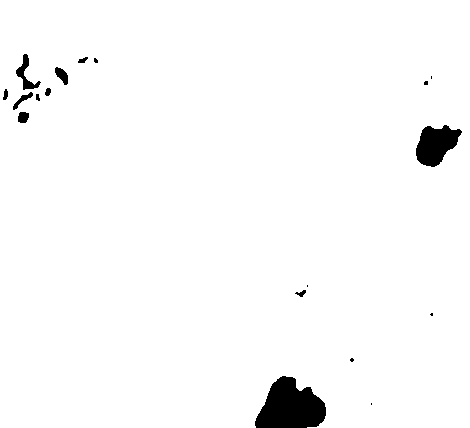}}
			\frame{\includegraphics[height=\sizemat]{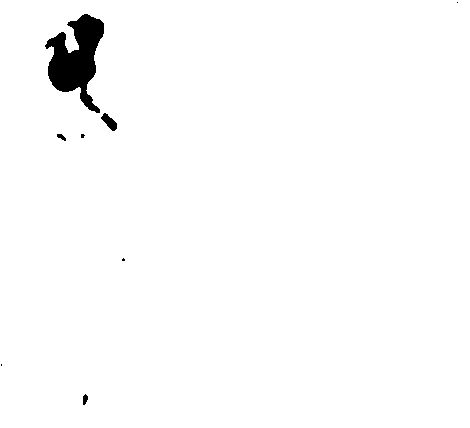}}
			\frame{\includegraphics[height=\sizemat]{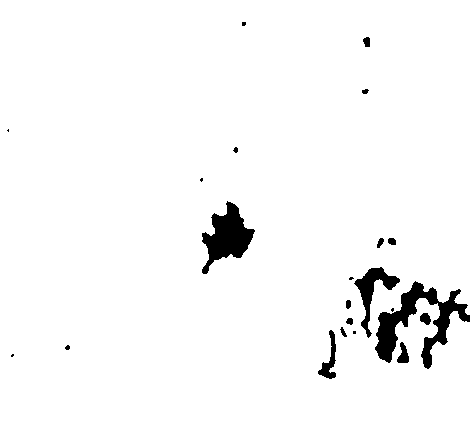}}
		}
		\caption{}\label{sbfig:original_images}
	\end{subfigure}
	
	\vspace{2ex}
		
	\begin{subfigure}{\textwidth}
		\centering
		\makebox[\textwidth]{	
			\frame{\includegraphics[height=\sizesgt]{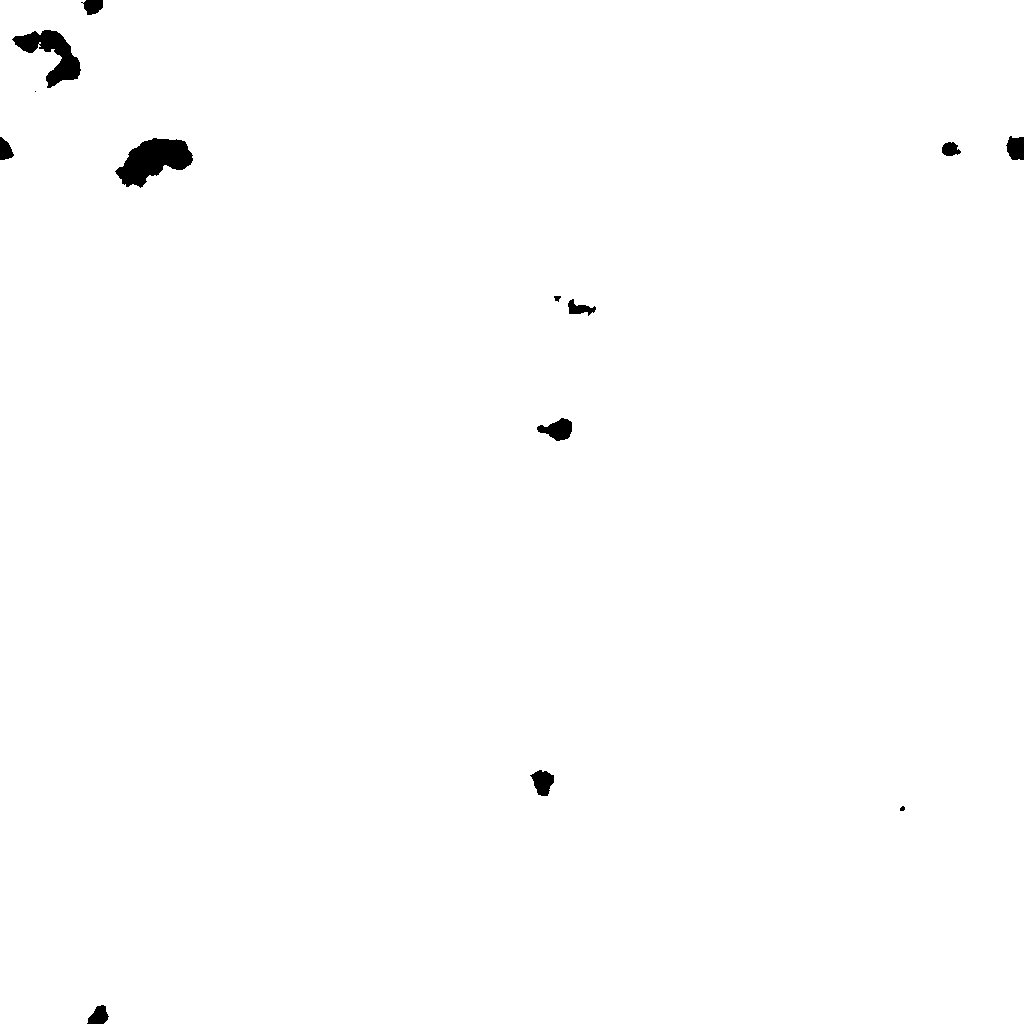}} 	
			\frame{\includegraphics[height=\sizesgt]{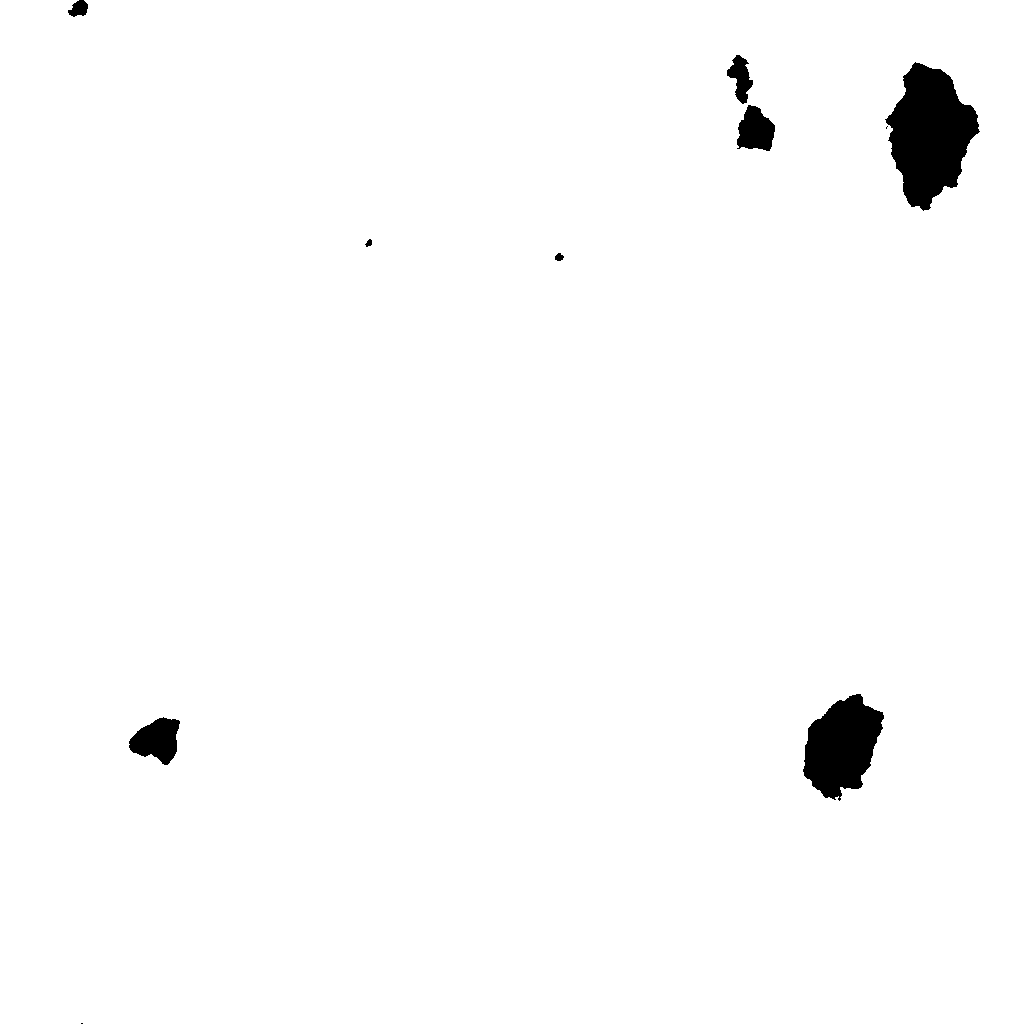}}
			\frame{\includegraphics[height=\sizesgt]{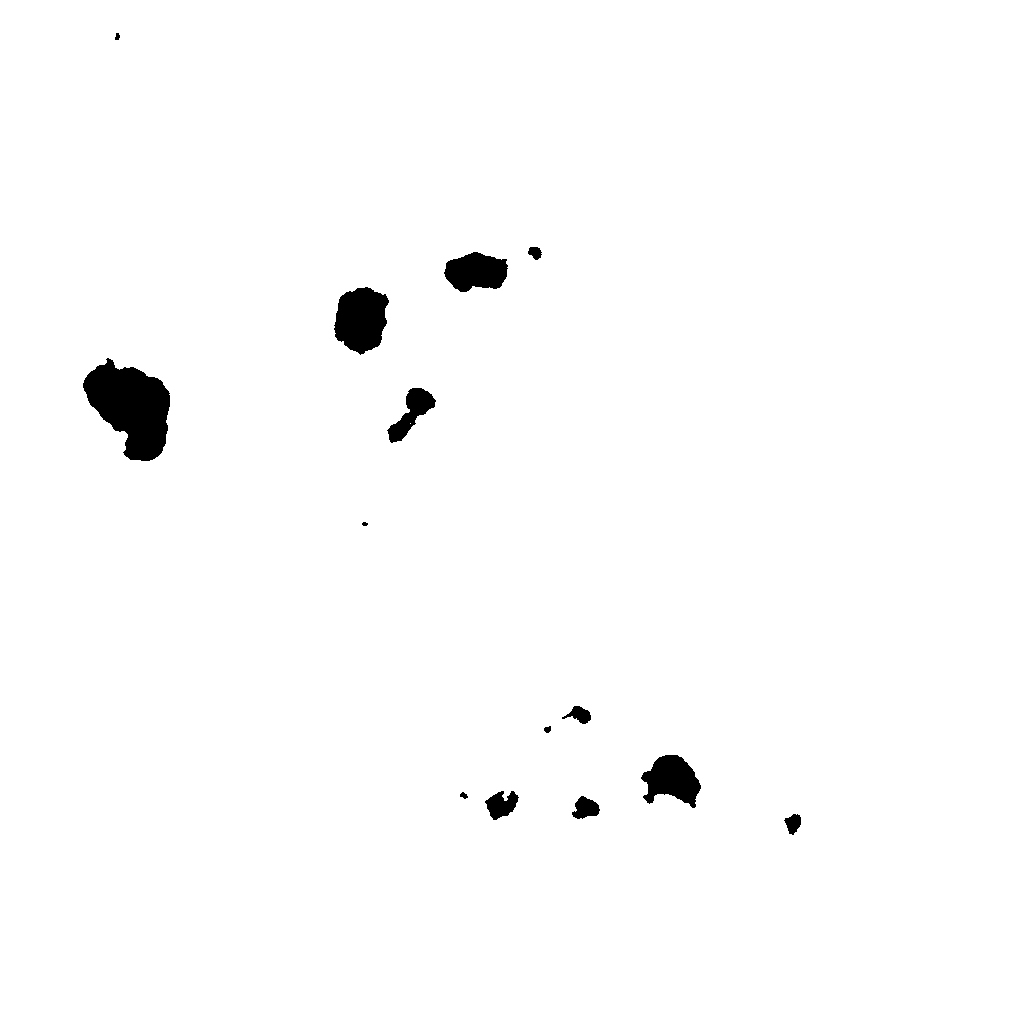}}
			\frame{\includegraphics[height=\sizesgt]{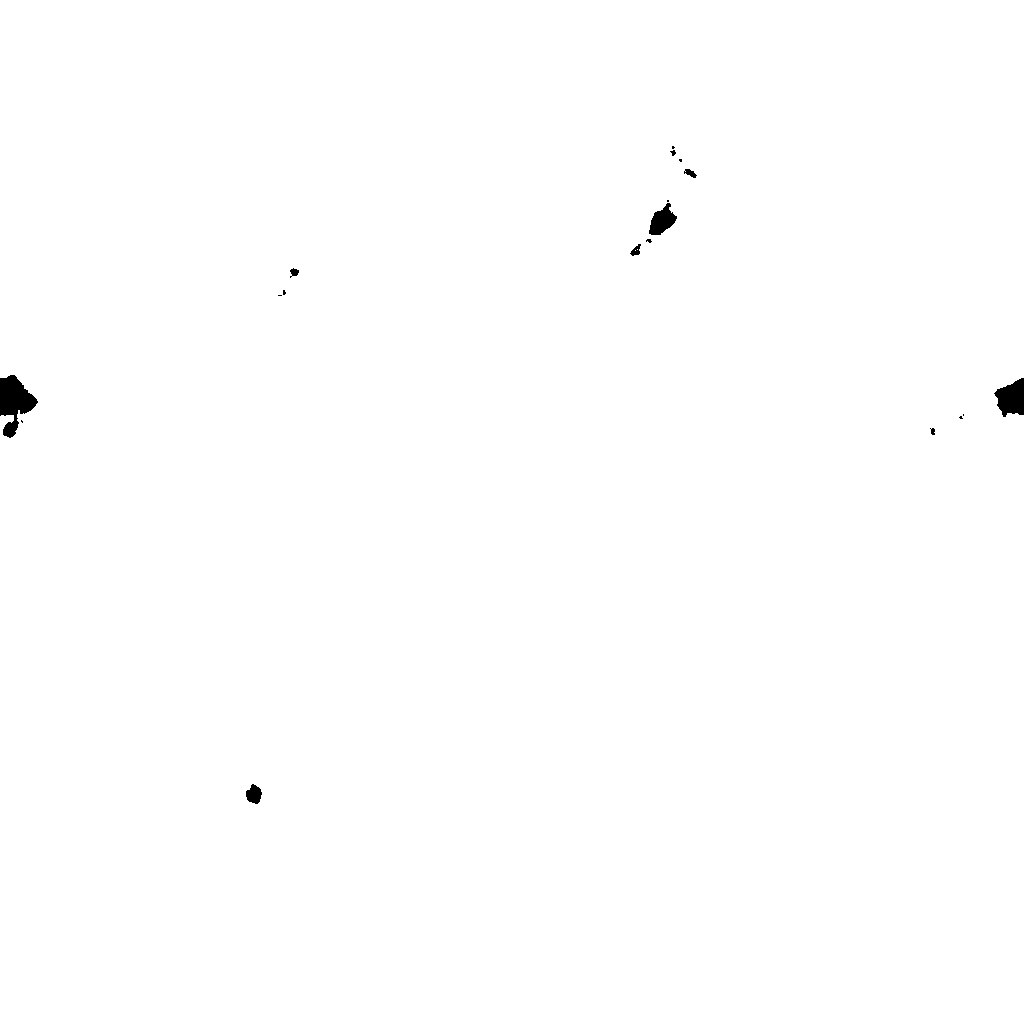}}
		}
		
		\vspace{0.5ex}				
		\makebox[\textwidth]{
			\frame{\includegraphics[height=\sizesgt]{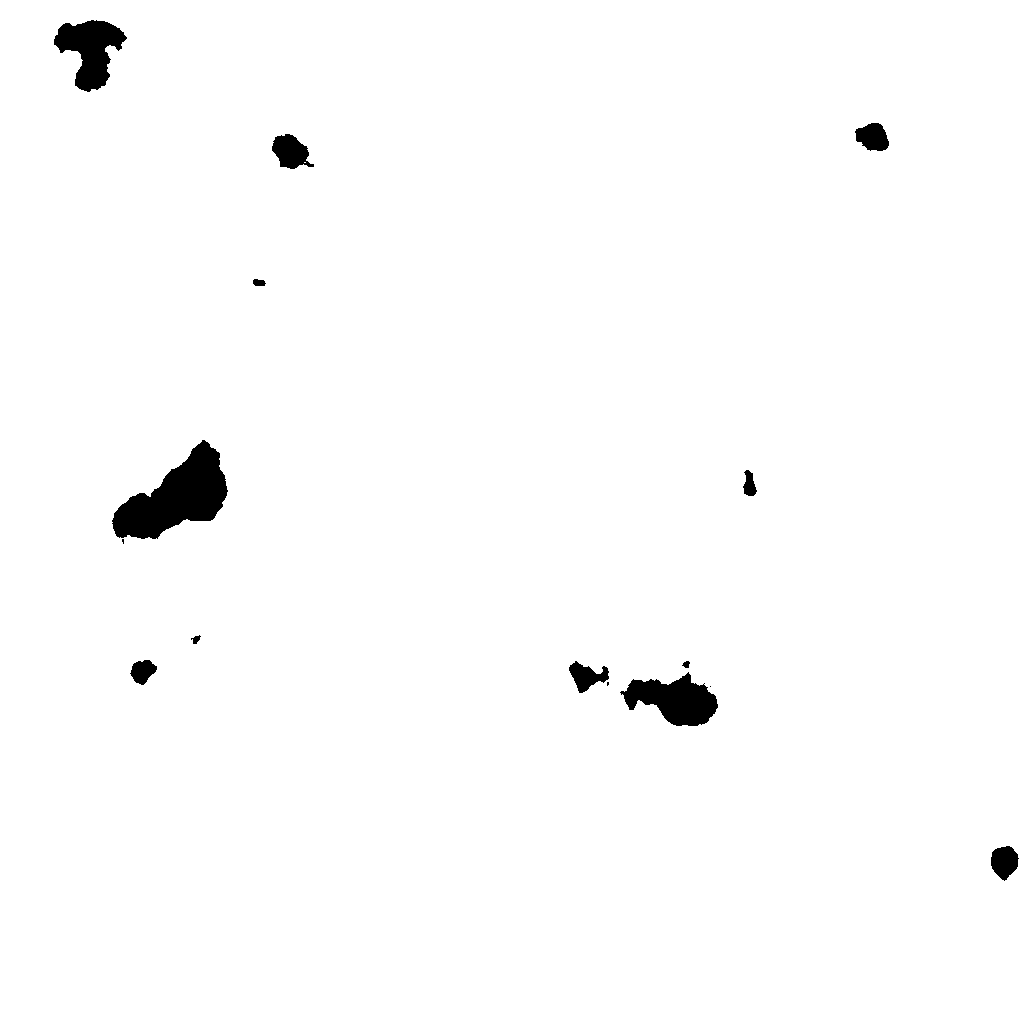}}
			\frame{\includegraphics[height=\sizesgt]{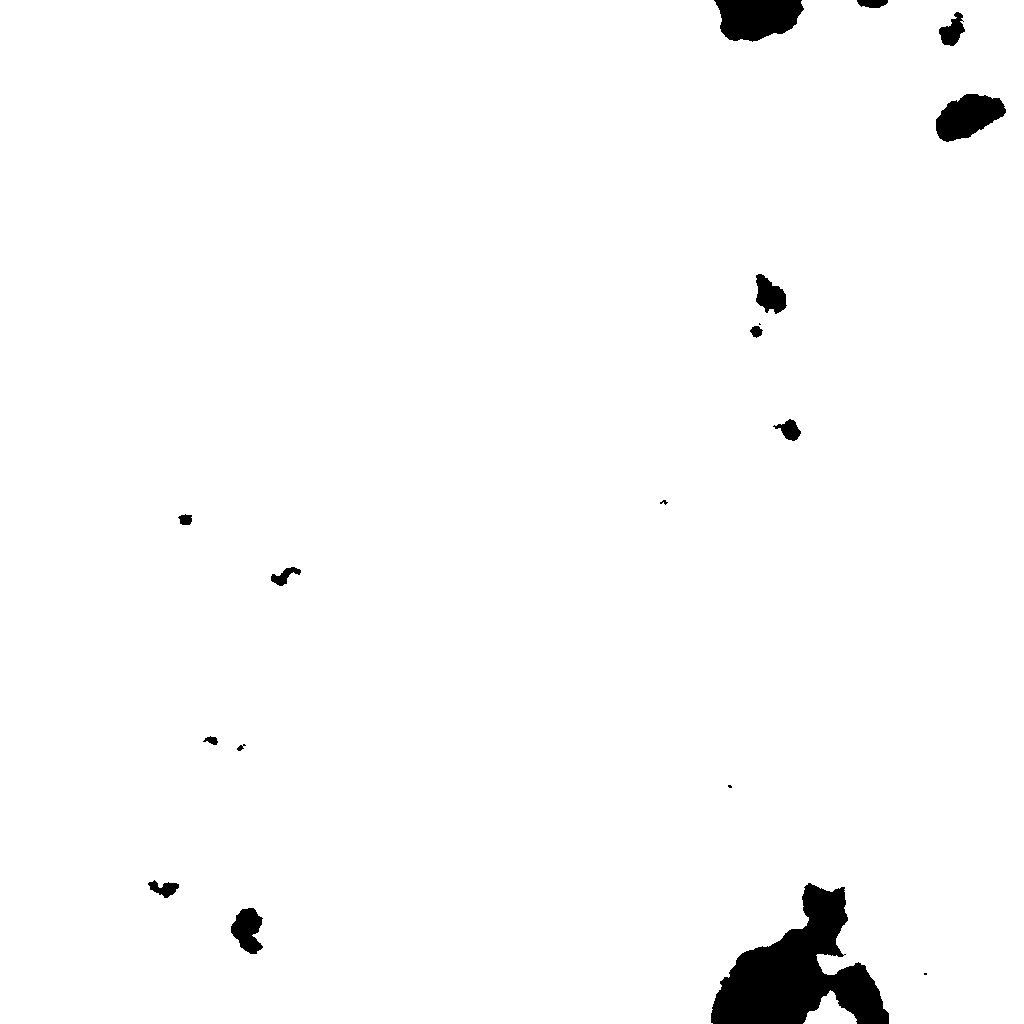}}
			\frame{\includegraphics[height=\sizesgt]{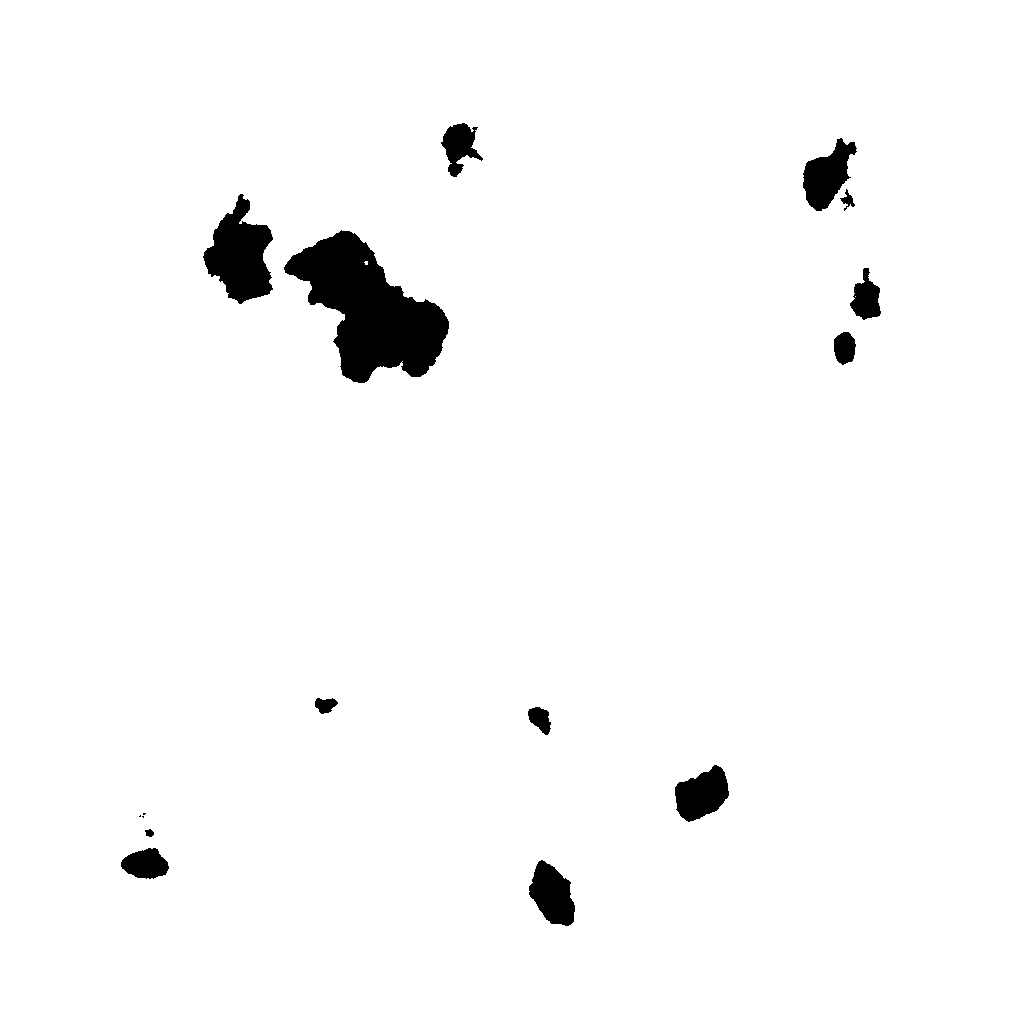}}
			\frame{\includegraphics[height=\sizesgt]{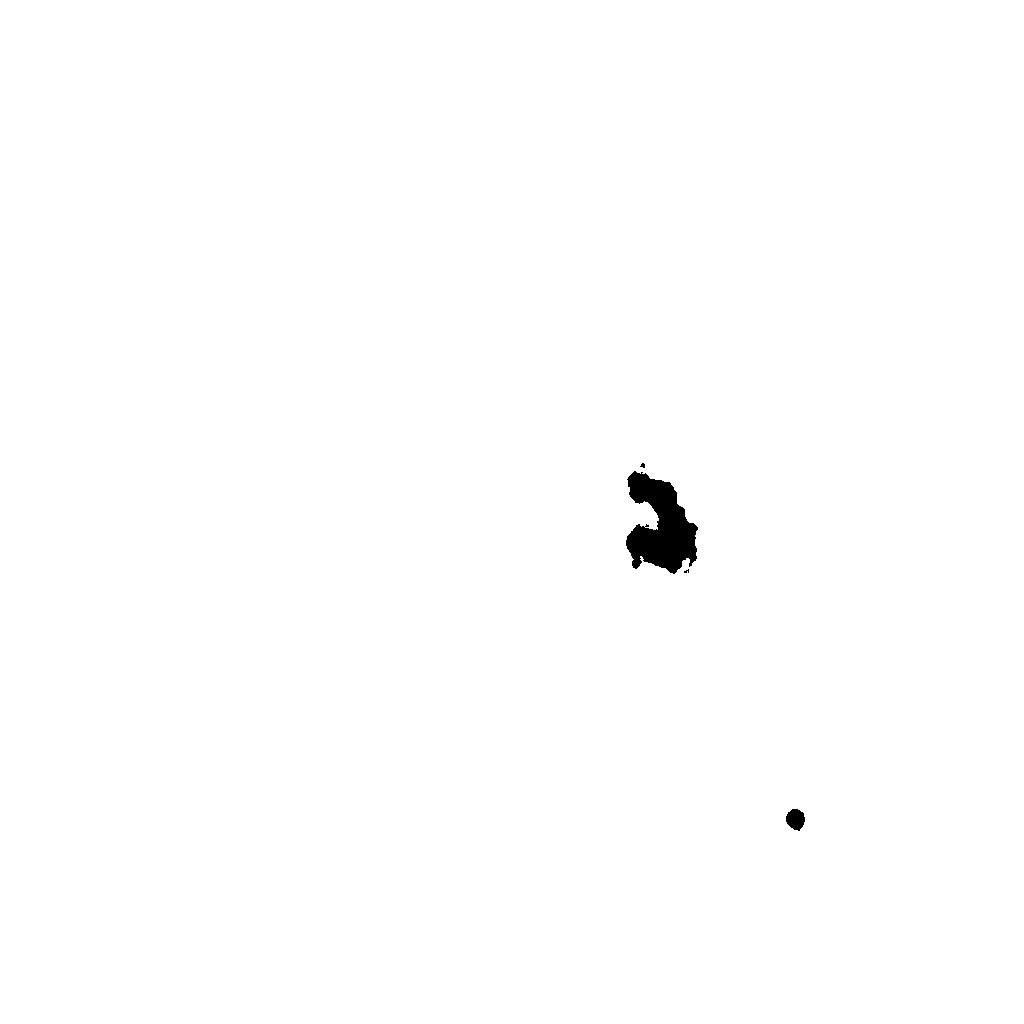}}
		}
		\caption{}\label{sbfig:surrogate_images_2D}
	\end{subfigure}

	\vspace{-2ex}
	\begin{subfigure}{\textwidth}
		\centering				
		\makebox[\textwidth]{
			\includegraphics[width=\sizesgtv]{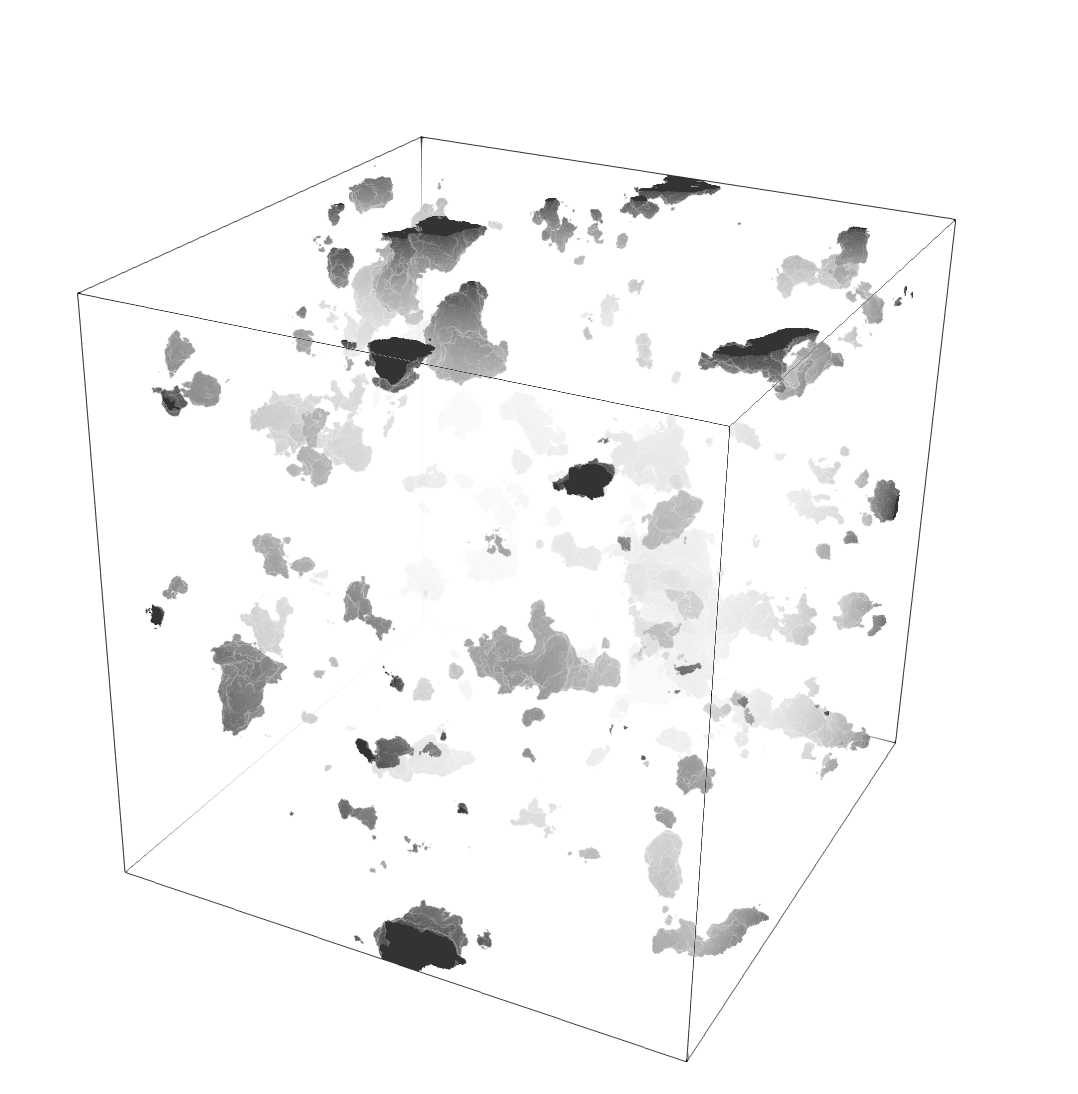}\hspace{5ex}	
			\includegraphics[width=\sizesgtv]{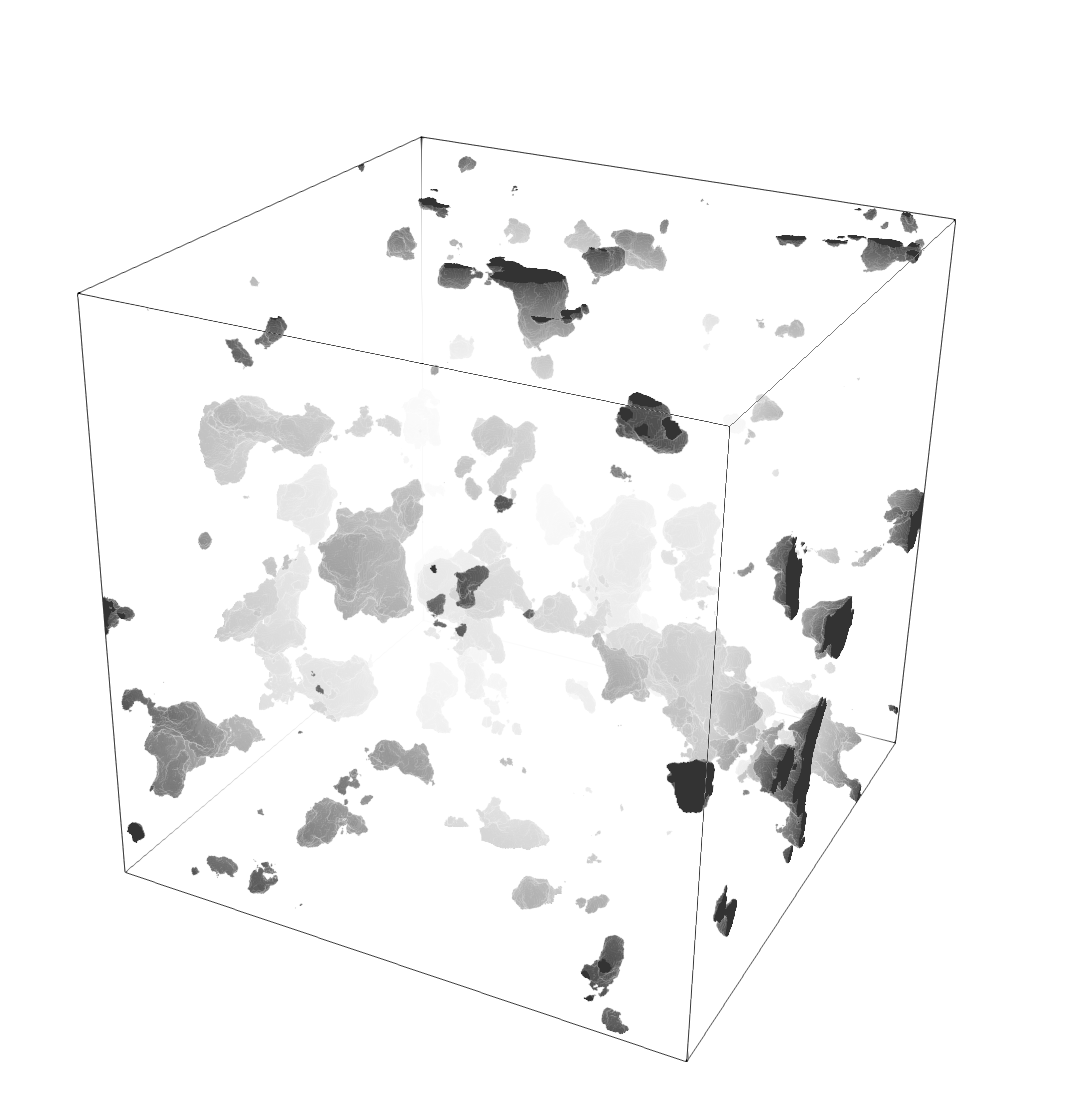}
		}
		\vspace{-5ex}
		\caption{}\label{sbfig:surrogate_images_3D}
	\end{subfigure}
	
	\caption{Original samples~(\subref{sbfig:original_images}) from~\cite{charkaluk2014probability} compared to the surrogate 2D~(\subref{sbfig:surrogate_images_2D}) and 3D~(\subref{sbfig:surrogate_images_3D}) samples. Design parameters are given by~\eqref{eq:reconstructed_bayesian}. Level of gray serves for 3D representation.}
	\label{fig:real_vs_surrogate}	
\end{figure}

\begin{figure}[!ht]
	\centering
	\begin{tikzpicture}[scale=1]
	
	\begin{groupplot}[
		axis standard,
		group style={group name=myplot,group size= 2 by 1, horizontal sep=2cm,},
		legend style={font=\scriptsize, at={(1.1,0.98)},anchor=north east}			
	]
	
	\nextgroupplot[
		ylabel={$S_2$},
		xlabel={$r$ (mm)},
		legend columns=3,
		legend style={at={(0.5,-0.4)},anchor=north,font=\scriptsize},
		ymax=0.028,
		ymin=0,
		xmin=0,
		xmax=0.5,
		cycle list/Paired,
		scaled y ticks = false,
	]
		\pgfplotstableread[header=true, col sep=comma]{data_files/plot_Cov.csv}{\data}
		\pgfplotstablegetcolsof{\data}\pgfmathsetmacro{\N}{\pgfplotsretval-2}
		
		\pgfplotsforeachungrouped \i in {1,...,\N} {
			\pgfmathtruncatemacro{\redfrac}{\i/\N*100}
			\edef\temp{
				\noexpand\addplot+[mark=none, thin, mark options={scale=0.5}] table[x index=0, y index=\i, header=true, col sep=comma] {\noexpand\data};
			}\temp
			\addlegendentryexpanded{ sample \i$\quad$}
		}
	
		\pgfmathsetmacro{\M}{\N+1}
		\addplot+[mark=none, very thick, color=black, mark options={scale=0.5}] table[x index=0, y index=\M, header=true, col sep=comma] {\data};
		\addlegendentry{ MLE }
	
	\nextgroupplot[
		ylabel={PDF},
		xlabel={Pore diameter (mm)},
		legend style={at={(0.5,-0.4)},anchor=north,font=\scriptsize},
		ymin=0,
		xmin=0,
		xmax=0.5,
	]
		\pgfplotstableread[header=true, col sep=comma]{data_files/plot_Rec_PoreSize.csv}{\data}	
		\addplot+[color=blue, thick, mark=None] table[x=binsref, y=Plog, header=true, col sep=comma] {\data};
		\addplot+[color=red, thick, mark=None] table[x=binsref, y=Pexp, header=true, col sep=comma] {\data};
		\addplot+[ybar, draw=none, color=black, bar width=1pt, fill=gray, very thin, mark=none, mark options={solid}, mark size=1pt, thin] table[x=bins, y=P, header=true, col sep=comma] {\data};
		\legend{Experimental (LogNormal), Experimental (Exponential), Surrogate}

	\end{groupplot}
	\end{tikzpicture}
	\vspace{-2ex}
	\caption{Left: the two-point correlation functions~$S_2(r)$ of the original samples from~\cite{charkaluk2014probability} (see Figure~\ref{sbfig:original_images}) and of the associated surrogate model (Minimal Likelihood estimator) with ${\vfm}_{\operatorname{MLE}}\approx 0.014$ and $\nu_{\operatorname{MLE}}=1.13$.
		Right: comparison of the poresize probability distribution of the surrogate material to the reference distributions from~\cite{charkaluk2014probability}.}
	\label{fig:PoresizeDist}
\end{figure}
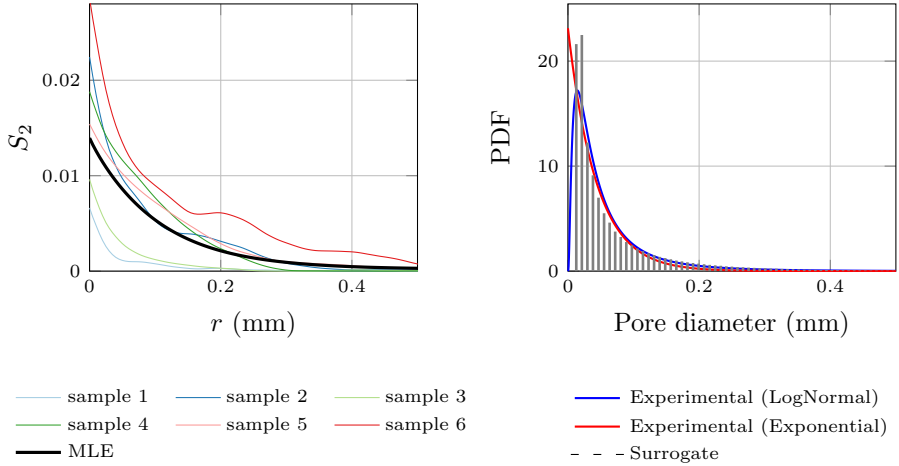

We reconstruct an aluminum–silicon alloy, A319-LFC, studied in~\cite{charkaluk2014probability}, using six 2D CT images of the alloy (Figure~\ref{fig:real_vs_surrogate}, top).
Using the Laplace approximation for Bayesian inference (see Section~\ref{sec:Reconstruction} for the procedure) of the design parameters of the surrogate material model introduced in Section~\ref{sec:Model}, we obtain
\begin{equation}\label{eq:reconstructed_bayesian}
	\mat{\log{\tau} \\ \log{\nu}} \sim \Gaussian\left(\mat{ \log{2.46}\\ \log{1.13}}, \mat{ 6.9\cdot 10^{-6} & 1.3\cdot 10^{-4} \\ 1.3\cdot 10^{-4} & 2.7\cdot 10^{-3}}\right).
\end{equation}
The estimator~$\tau_{\operatorname{MLE}}$ corresponds to porosity~${\vfm}_{\operatorname{MLE}}\approx 0.014$, $\nu$ varies in~$(1.0,\,1.26)$.
The correlation length is $\corrlen\approx 0.256\,mm$.
With the design parameters in hand, we can generate the surrogate microstructure samples in 2D or in 3D.
The original CT images and the surrogate 2D and 3D samples are presented in Figure~\ref{fig:real_vs_surrogate}.
The two-point correlation function~$S_2(r)$ obtained with MLE parameters is traced in Figure~\ref{fig:PoresizeDist} (left) and compared with two-point correlations of original images (computed as cross-correlations using FFT).
In~\cite{charkaluk2014probability}, the experimentally computed distribution has been fitted with a lognormal and exponential distributions. 
In Figure~\ref{fig:PoresizeDist} (right), we compare these reference distributions to the surrogate pore-size distribution, obtained from $10,000$ 3D samples of $2^{8\cdot 3}\approx {1.6\cdot 10^{7}}$ voxels.
We observe that both distributions are in very good agreement.

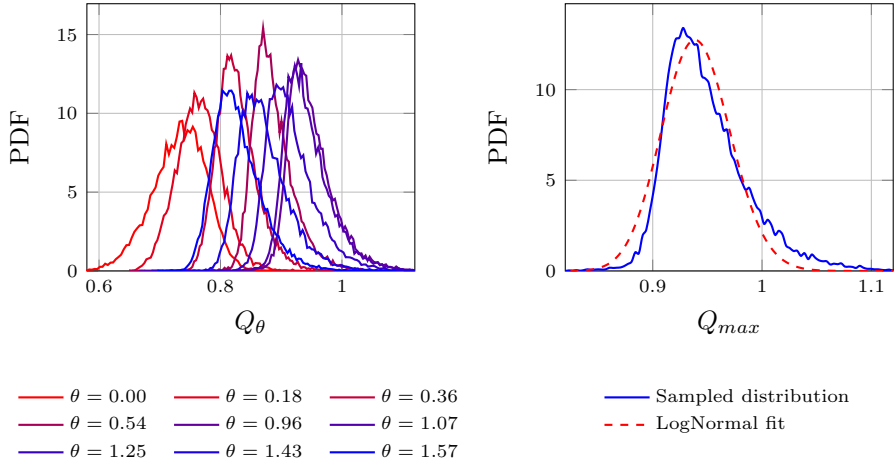
\begin{figure}[!ht]
	\centering
	
	\begin{tikzpicture}[scale=1.0]	
		
		\begin{groupplot}[
			axis standard,
			group style={group name=myplot,group size= 2 by 1, horizontal sep=2cm,},			
		]
		
		\nextgroupplot[
			ylabel={PDF},
			xlabel={$\QoI_{\theta}$},
			legend columns=3,
			legend style={at={(0.5,-0.4)},anchor=north,font=\scriptsize},
			xmax=1.12,
			xmin=0.58,
			ymin=0,
		]
		
			\pgfplotstableread[header=true, col sep=comma]{data_files/plot_Rec_Q.csv}{\data}			
			
			\pgfplotstablegetcolsof{\data}\pgfmathsetmacro{\N}{int(\pgfplotsretval/2)-1}
			
			\pgfplotsforeachungrouped \i in {0,5,10,15,27,30,35,40,\N} {
				\pgfmathtruncatemacro{\xindex}{int(2*\i)}
				\pgfmathtruncatemacro{\yindex}{int(2*\i+1)}
				\pgfmathtruncatemacro{\redfrac}{\i/\N*100}
				\edef\temp{
					\noexpand\addplot[mark=none, color=blue!\redfrac!red, mark options={scale=0.5}, thick] table[x index=\xindex, y index=\yindex, header=true, col sep=comma] {\noexpand\data};
				}\temp
				\pgfplotstablegetcolumnnamebyindex{\xindex}\of{\data}\to\thetaval
				\addlegendentryexpanded{ $\theta=\thetaval\quad$ }
			}

		\nextgroupplot[
			ylabel={PDF},
			xlabel={$\QoI_{max}$},
			legend style={at={(0.5,-0.4)},anchor=north,font=\scriptsize},
			xmax=1.12,
			xmin=0.82,
			ymin=0,
		]
			\pgfplotstableread[header=true, col sep=comma]{data_files/plot_Rec_Qmax.csv}{\data}
			\addplot+[color=blue, thick, smooth, mark=none] table[x=bins, y=P, header=true, col sep=comma] {\data};
			\addplot+[color=red, thick, dashed, smooth, mark=none] table[x=bins, y=Pfit, header=true, col sep=comma] {\data};
			\legend{Sampled distribution, LogNormal fit}		
		\end{groupplot}
	\end{tikzpicture}
	
	\caption{Probability distribution of~$\QoI_{\theta}$ for different loads~$\theta$ (left) and for the "worst" loading~$\theta_{\max}\approx 0.3\,\pi$ (right), which maximizes the average of~$\QoI_{\theta}$.}
	\label{fig:surrogate_dist}
\end{figure}

For the surrogate material, we compute the quantity of interest~$\QoI(\theta)$ for different loading types~$\theta$, using $20,000$ 3D samples of $2^{7\cdot 3}\approx {2\cdot 10^{6}}$ voxels.
The angle~$\theta$ is discretized with $45$ equidistant points in $[0,\pi/2]$.
The probability distributions of the quantity of interest~$\QoI(\theta)$ are shown in Figure~\ref{fig:surrogate_dist}~(left).
The "worst" loading case~$\theta_{\max}$, maximizing the average of~$\QoI(\theta)$, is given by~$\theta_{\max}\approx 0.3\,\pi$.
The associated distribution is also depicted in Figure~\ref{fig:surrogate_dist}~(right), along with its lognormal fit.

\subsection{Sensitivity to the design parameters}
\label{sec:AppSensitivity}

\begin{figure}[!ht]
	\centering
	\includegraphics[width=\textwidth]{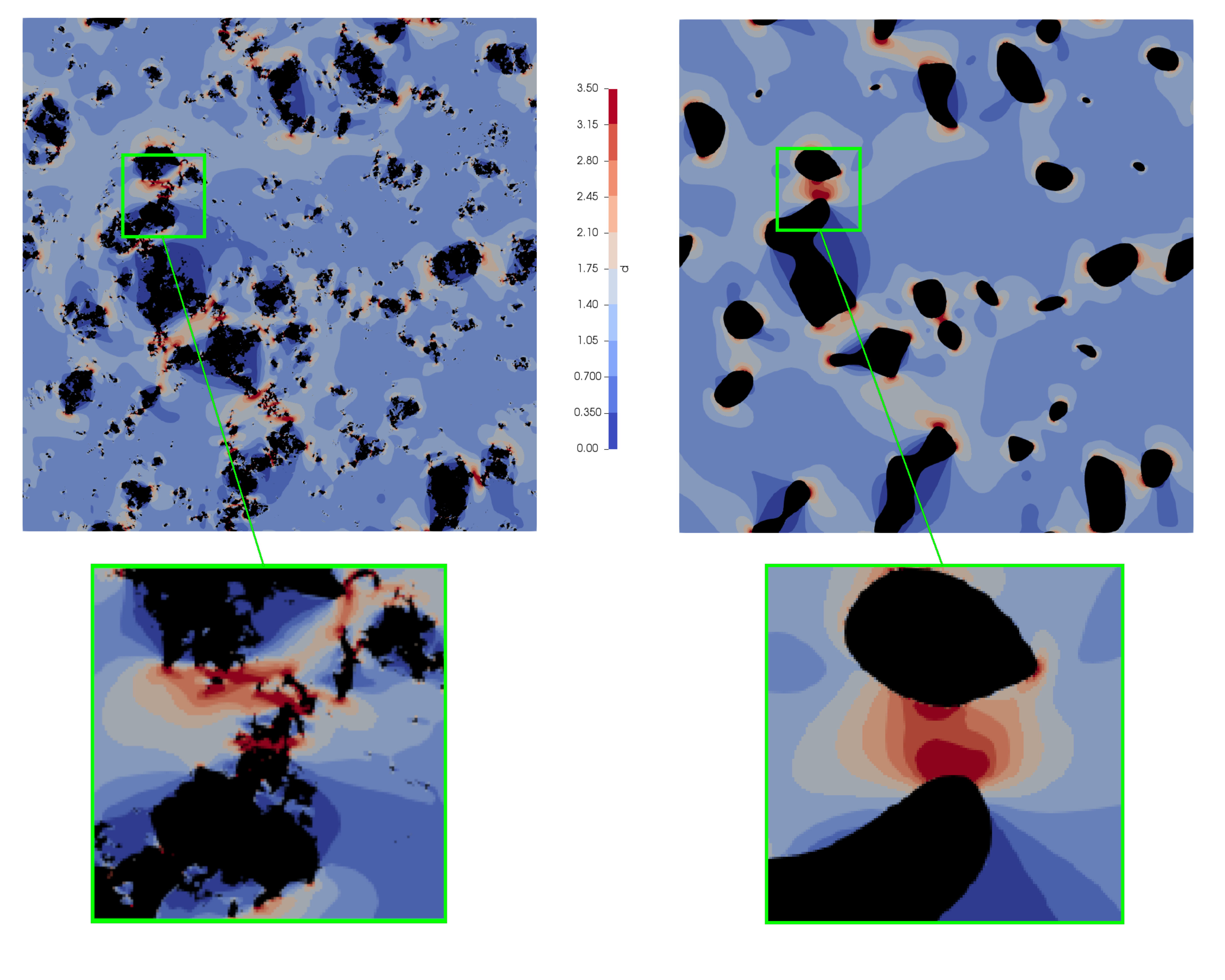}
	\vspace{-8ex}
	\caption{Examples of solution: damage parameter field~$d(\x)$~(\ref{eq:damage}) for $\nu=0.5$~(left) and $\nu=10$~(right) with porosity~$\vfm=0.1$, zoom of the zone of the maximum damage parameter. }
	\label{fig:solution}
\end{figure}

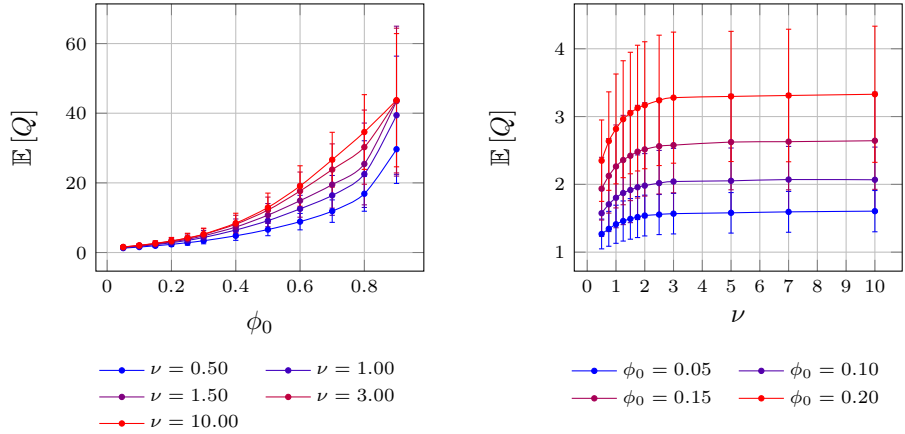
\begin{figure}[!ht]
	\centering
	
	\begin{tikzpicture}[scale=1.0]

	\begin{groupplot}[
	axis standard,
	group style={group name=myplot,group size= 2 by 1, horizontal sep=2cm,},
	]
	
	\nextgroupplot[
	ylabel={$\E{\QoI}$},
	xlabel={$\vfm$},
	legend columns=2,
	legend style={at={(0.5,-0.3)},anchor=north,font=\scriptsize},
	]
	
	\pgfplotstableread[header=true, col sep=comma]{data_files/plot_Q.csv}{\data}
	\pgfplotstablegetcolsof{\data}\pgfmathsetmacro{\Nnu}{int((\pgfplotsretval-1)/2)}			
	

	\pgfplotsforeachungrouped \i in {1,...,\Nnu} {
		\pgfmathtruncatemacro{\yindex}{int(2*\i-1)}
		\pgfmathtruncatemacro{\yerrorindex}{int(2*\i)}
		\pgfmathtruncatemacro{\redfrac}{(\i-1)*100/(\Nnu-1)}
		\edef\temp{
			\noexpand\addplot[line standard, thin, color=red!\redfrac!blue, enable error bars] table[x index=0, y index=\yindex, y error index=\yerrorindex, header=true, col sep=comma] {\noexpand\data};
		}\temp
		\pgfplotstablegetcolumnnamebyindex{\yindex}\of{\data}\to\nuval
		\addlegendentryexpanded{ $\nu=\nuval\quad$ }
	}
	
	\nextgroupplot[
	ylabel={$\E{\QoI}$},
	xlabel={$\nu$},
	legend columns=2,
	legend style={at={(0.5,-0.3)},anchor=north,font=\scriptsize},
	xtick={0,1,2,3,4,5,6,7,8,9,10},
	]
	
	\pgfplotstableread[header=true, col sep=comma]{data_files/plot_Q_nu.csv}{\data}	
	\pgfplotstablegetcolsof{\data}\pgfmathsetmacro{\Nvf}{int((\pgfplotsretval-1)/2)}
	
	
	\pgfplotsforeachungrouped \i in {1,...,\Nvf} {
		\pgfmathtruncatemacro{\yindex}{int(2*\i-1)}
		\pgfmathtruncatemacro{\yerrorindex}{int(2*\i)}
		\pgfmathtruncatemacro{\redfrac}{(\i-1)*100/(\Nvf-1)}
		\edef\temp{
			\noexpand\addplot[line standard, thin, color=red!\redfrac!blue, enable error bars] table[x index=0, y index=\yindex, y error index=\yerrorindex, header=true, col sep=comma] {\noexpand\data};
		}\temp
		\pgfplotstablegetcolumnnamebyindex{\yindex}\of{\data}\to\vfval
		\addlegendentryexpanded{ $\vfm=\vfval\quad$ }
	}
	
	\end{groupplot}
	\end{tikzpicture}
	
	\vspace{-2ex}
	\caption{Average of $\QoI$ as function of porosity~$\vfm$ (left) and of pore regularity~$\nu$ (right).}
	\label{fig:design_vf}
\end{figure}

In what follows, $\QoI$ is associated with the load $\theta_{\max}=\max_{0\le\theta\le\frac{\pi}{2}}\E{\QoI(\theta)}$, which corresponds to~$0.3\,\pi<\theta_{\max}<0.32\,\pi$.
We want to study the sensitivity of the quantity of interest~$\QoI$ (\ref{eq:QoI}) to the two design parameters: porosity~$\vfm$ and pore regularity~$\nu$.
To be representative, a volume element has to be of size much larger than the correlation length~$\corrlen$.
So we consider a unit square as RVE and fix $\corrlen=0.05$.
For each $\vfm$ and $\nu$, we solve the plane-strain problem~(\ref{eq:LE in strains}) on $10,000$ 2D samples with $2^{8\cdot2}\approx 6.5\cdot 10^4$~voxels.
Young's modulus is $1\,GPa$ for the material matrix and $0$ in the pores, material Poisson ratio is $0.3$.
Solution examples for $\nu=0.5$ and $\nu=10$ with porosity~$\vfm=0.1$ are shown in Figure~\ref{fig:solution}. 
Figure~\ref{fig:design_vf} depicts the average of~$\QoI$ as function of porosity~$\vfm$ and of pore regularity~$\nu$.
The standard deviation is denoted with the error bars.
We observe that the damage parameter grows with the porosity and with the pore regularity.
We note the asymptotic behavior of~$\QoI$ when $\nu$ goes to infinity.
This limit case corresponds to infinitely differentiable pore interfaces, when the Mat\'ern covariance of the intensity field becomes a squared exponential covariance.
We also note that the influence of the parameter~$\nu$ increases with the porosity.
The standard deviation of~$\QoI$ also grows with the porosity.

\begin{figure}[!ht]
	\centering
	
	\begin{tikzpicture}[scale=1.0]

		\begin{groupplot}[
			axis standard,
			group style={group name=myplot,group size= 2 by 1, horizontal sep=2cm,},
			cycle list/Paired,	
		]
		
			\nextgroupplot[
				ylabel={$\bulk$},
				xlabel={$\vfm$},
				legend columns=6,
				legend style={at={(1.25,-0.4)},anchor=north,font=\scriptsize},
			]
			
				\pgfplotstableread[header=true, col sep=comma]{data_files/plot_K.csv}{\data}
				\pgfplotstablegetcolsof{\data}\pgfmathsetmacro{\Nnu}{int((\pgfplotsretval-3)/2)}				
				

				\addplot+[line standard, color=black, very thick] table[x index=0, y index=2,  header=true, col sep=comma] {\data};
				\addlegendentry{ HS $\quad$ }

				\pgfplotsforeachungrouped \i in {1,...,\Nnu} {
					\pgfmathtruncatemacro{\yindex}{int(2*\i+1)}
					\pgfmathtruncatemacro{\yerrorindex}{int(2*\i+2)}
					\pgfmathtruncatemacro{\redfrac}{(\i-1)*100/(\Nnu-1)}
					\edef\temp{
						\noexpand\addplot+[line standard, thin, enable error bars] table[x index=0, y index=\yindex, y error index=\yerrorindex, header=true, col sep=comma] {\noexpand\data};
					}\temp
					\pgfplotstablegetcolumnnamebyindex{\yindex}\of{\data}\to\nuval
					\addlegendentryexpanded{ $\nu=\nuval\quad$ }
				}

			\nextgroupplot[
				ylabel={$\shear$},
				xlabel={$\vfm$},
			]			
			
				\pgfplotstableread[header=true, col sep=comma]{data_files/plot_G.csv}{\data}
				\pgfplotstablegetcolsof{\data}\pgfmathsetmacro{\Nnu}{int((\pgfplotsretval-3)/2)}				
				
			
				\addplot+[line standard, color=black, very thick] table[x index=0, y index=2,  header=true, col sep=comma] {\data};
			
				\pgfplotsforeachungrouped \i in {1,...,\Nnu} {
					\pgfmathtruncatemacro{\yindex}{int(2*\i+1)}
					\pgfmathtruncatemacro{\yerrorindex}{int(2*\i+2)}
					\pgfmathtruncatemacro{\redfrac}{(\i-1)*100/(\Nnu-1)}
					\edef\temp{
						\noexpand\addplot+[line standard, thin, enable error bars] table[x index=0, y index=\yindex, y error index=\yerrorindex, header=true, col sep=comma] {\noexpand\data};
					}\temp
				}
			
		\end{groupplot}
	\end{tikzpicture}
	
	\caption{Homogenized elastic moduli $\bulk$ (bulk) and $\shear$ (shear) as functions of porosity~$\vfm$ for different values of~$\nu$, compared to the corresponding upper Hashin-Shtrikman bounds (HS).}
	\label{fig:HS}
\end{figure}
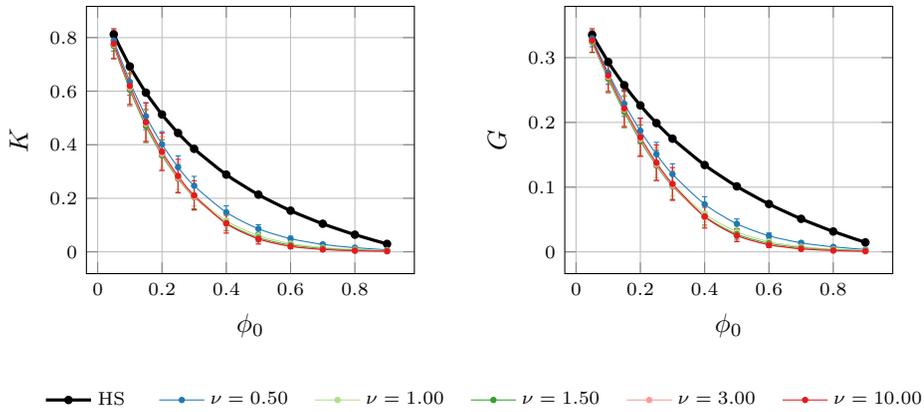

Finally, in addition, we want to estimate the homogenized elastic moduli, $\bulk$ (bulk) and $\shear$ (shear). 
Figure~\ref{fig:HS} shows their averages as functions of porosity, compared to the corresponding upper Hashin-Shtrikman bounds~\cite{hashin1963}.
The standard deviation is presented there with error bars.
As before, we observe the standard deviation growing with the porosity.

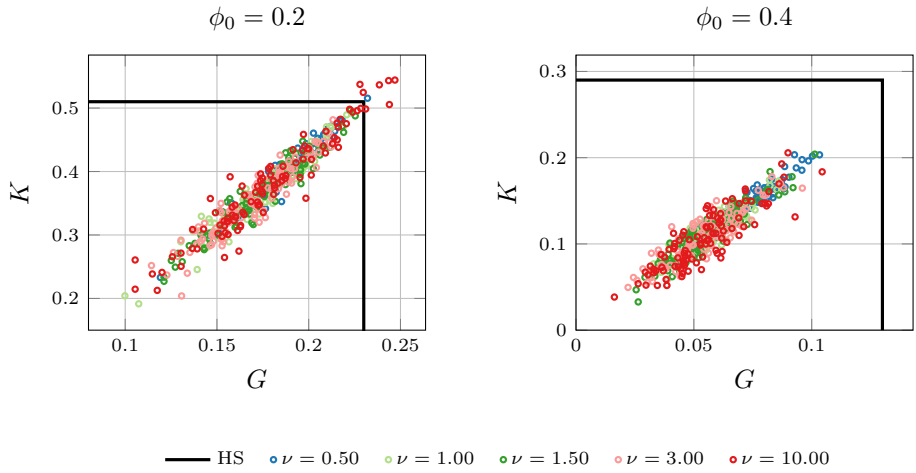
\begin{figure}[!ht]
	\centering
	
	\begin{tikzpicture}[scale=1.0]
	
	\begin{groupplot}[
		group style={group name=myplot,group size= 2 by 1, horizontal sep=2cm,},
		axis standard,
		width=0.5\textwidth,
		ylabel={$\bulk$},
		xlabel={$\shear$},
		cycle list name=black white,
		cycle list/Paired,	
	]
	
		\nextgroupplot[
			legend columns=6,
			legend style={at={(1.25,-0.4)},anchor=north,font=\scriptsize},
			xmin=0.08,
			ymin=0.15,
			title={$\vfm=0.2$},
		]
		
			\pgfplotstableread[header=true, col sep=comma]{data_files/plot_KGscatter.csv}{\data}
			\pgfplotstablegetcolsof{\data}\pgfmathsetmacro{\Nnu}{int((\pgfplotsretval-4)/2)}
			
			\pgfplotstablegetcolumnnamebyindex{0}\of{\data}\to{\HSlowK}
			\pgfplotstablegetcolumnnamebyindex{1}\of{\data}\to{\HSupK}
			\pgfplotstablegetcolumnnamebyindex{2}\of{\data}\to{\HSlowG}
			\pgfplotstablegetcolumnnamebyindex{3}\of{\data}\to{\HSupG}
			\addplot+[mark=none, very thick, black] coordinates { (\HSlowG,\HSlowK) (\HSlowG,\HSupK) (\HSupG,\HSupK) (\HSupG,\HSlowK) (\HSlowG,\HSlowK) };
			\addlegendentry{ HS$\quad$ }

			\pgfplotsforeachungrouped \i in {1,...,\Nnu} {
				\pgfmathtruncatemacro{\Kindex}{int(2*\i+2)}
				\pgfmathtruncatemacro{\Gindex}{int(2*\i+3)}
				\pgfmathtruncatemacro{\redfrac}{(\i-1)*100/(\Nnu-1)}
				\edef\temp{
					\noexpand\addplot+[mark=o, only marks, mark options={scale=0.5}, thick] table[x index=\Gindex, y index=\Kindex, header=true, col sep=comma] {\noexpand\data};
				}\temp
				\pgfplotstablegetcolumnnamebyindex{\Kindex}\of{\data}\to\nuval
				\addlegendentryexpanded{ $\nu=\nuval\quad$ }
			}

		\nextgroupplot[
			xmin=0.00,
			ymin=0.00,
			title={$\vfm=0.4$},
		]			
		
			\pgfplotstableread[header=true, col sep=comma]{data_files/plot_KGscatter_vf=0.40.csv}{\data}
			\pgfplotstablegetcolsof{\data}\pgfmathsetmacro{\Nnu}{int((\pgfplotsretval-4)/2)}
			
			\pgfplotstablegetcolumnnamebyindex{0}\of{\data}\to{\HSlowK}
			\pgfplotstablegetcolumnnamebyindex{1}\of{\data}\to{\HSupK}
			\pgfplotstablegetcolumnnamebyindex{2}\of{\data}\to{\HSlowG}\pgfmathsetmacro{\HSlowG}{0}			
			\pgfplotstablegetcolumnnamebyindex{3}\of{\data}\to{\HSupG}
			\addplot+[mark=none, very thick, black] coordinates { (\HSlowG,\HSupK) (\HSupG,\HSupK) (\HSupG,\HSlowK) };

			\pgfplotsforeachungrouped \i in {1,...,\Nnu} {
				\pgfmathtruncatemacro{\Kindex}{int(2*\i+2)}
				\pgfmathtruncatemacro{\Gindex}{int(2*\i+3)}
				\pgfmathtruncatemacro{\redfrac}{(\i-1)*100/(\Nnu-1)}
				\edef\temp{
					\noexpand\addplot+[mark=o, only marks, mark options={scale=0.5}, thick] table[x index=\Gindex, y index=\Kindex, header=true, col sep=comma] {\noexpand\data};
				}\temp
				\pgfplotstablegetcolumnnamebyindex{\Kindex}\of{\data}\to\nuval
			}
	
	\end{groupplot}
	
	\end{tikzpicture}
	
	\caption{Scattering of the elastic moduli $\bulk$ (bulk) and $\shear$ (shear) over $100$ samples for different values of~$\nu$ (marked by color) and fixed porosity $\vfm=0.2$ (left) and $\vfm=0.4$ (right).}
	\label{fig:KGscatter}
\end{figure}

Figure~\ref{fig:KGscatter} shows the scattering of the homogenized moduli over $100$~samples for each value of~$\nu$ and fixed~$\vfm=0.2$.
We remark that the parameter~$\vfm$ is the average volume fraction, and thus the computed average $\bulk$ and $\shear$ respect the bound.
However, a particular realization has random volume fraction and can, therefore, violate the bound (see Figure~\ref{fig:KGscatter}).
For each sample, the associated homogenized moduli are approximated as
\begin{equation}
	\bulk = \frac{1}{d}\cdot\frac{\tr\Macrostress_{\theta=0}}{\tr\Macrostrain_{\theta=0}}, \qquad \shear = \frac{1}{2}\cdot\frac{\Macrostress_{\theta=\pi/2}}{\Macrostrain_{\theta=\pi/2}},
\end{equation}
where $\Macrostress_{\theta}$ is the imposed macro stress, and $\Macrostrain_{\theta}$ is the associated computed macro strain (see Subsection~\ref{sec:QoI}).

	\section{Conclusion}\label{sec:Conclusion}

In this work, we have demonstrated a framework for uncertainty quantification in fatigue analysis using a surrogate microstructure model.
The proposed microstructure model presents a trade off between computational performance, realistic random microstructure reconstruction and the small number of design parameters.

The approach presented here is quite general, although for specifity we have focused on the simple case of statistically isotropic media.
The anisotropy can be included through the shape operator~$\Shape$ in~(\ref{eq:MaternCovariance}).
Modifying the ratio of different correlation lengths, we change the pore aspect ratio.
In more sophisticated cases, when $\Shape(\x)$ depends on the position, the sampling procedure can be still based on the solution of the SPDE~(\ref{eq:SPDE}), however, the FFT can not be employed any more.

In this work, we have considered covariances of Matérn class~\cite{whittle1954stationary,matern1986spatial,handcock1993bayesian}, which presents a large class of covariance functions and provides enough flexibility in material reconstruction.
In addition to the traditional morphology descriptors as porosity, size and aspect ratio, this covariance class provides the regularity of the inclusion interface, which is also related to the pore sphericity~\cite{buffiere2001experimental,le2017}.
In order to introduce more design parameters and thus extend the class of possible reconstructed materials, one can consider other more general covariance kernels~\cite{matern1986spatial,stein2012interpolation,lim2009generalized,teubner1991level}.

We also demonstrated the use of the model in the fatigue analysis, and we have studied the influence of the regularity of the pore interface on the statistical properties of the simplified fatigue criteria and of the homogenized elastic moduli.


	\appendix

\section{Proof of Lemma~\ref{lem:moments}}
\label{apx:Moments}


\begin{proof}
	We show the formulas \eqref{eq:S1}-\eqref{eq:S2} for $S_1$ and $S_2$ in two steps.
	First, we consider $S_1$.
	Since the intensity~$\Int(\x;\omega)$ is a Gaussian random field, the mean of the phase~$\Phase(\x;\omega)$, defined by~\eqref{eq:levelcut}, satisfies 
	\begin{align}
		S_1 &= \frac{1}{\sqrt{2\pi\sigma^2}}\int\limits_{-\infty}^{\infty}\chi_{[\abs{\xi}>\tau]}\,\exp{-\frac{\xi^2}{2\sigma^2}} \d\xi \\
		&= \frac{1}{\sqrt{2\pi}}\int\limits_{\tau/\sigma}^{\infty}\exp{-\frac{1}{2}\xi^2} \d\xi + \frac{1}{\sqrt{2\pi}}\int\limits_{-\infty}^{-\tau/\sigma}\exp{-\frac{1}{2}\xi^2} \d\xi
		= \frac{2}{\sqrt{2\pi}}\int\limits_{\tau/\sigma}^{\infty}\exp{-\frac{1}{2}\xi^2} \d\xi
	\end{align}	
	Let us consider the volume fraction of all inclusions
	\begin{equation}
		\vf(\omega) = \frac{1}{\abs{D}}\int\limits_{D}\chi(\x;\omega)\d\x.
	\end{equation}
	Then, its average over samples~$\vfm=\E{\varphi}$ is
	\begin{align}
		\vfm = \E{\vf(\omega)}
		&= \frac{1}{\abs{D}}\int\limits_{D}\E{\chi(\x; \omega)}\d\x = \E{\chi} = P(\abs{\Int} > \tau) \\
		&= \frac{1}{\sqrt{2\pi\sigma^2}}\int\limits_{\tau}^{\infty} \exp{-\frac{\xi^2}{2\sigma^2}}\d\xi + \frac{1}{\sqrt{2\pi\sigma^2}}\int\limits_{-\infty}^{-\tau} \exp{-\frac{\xi^2}{2\sigma^2}}\d\xi 
		= \frac{2}{\sqrt{\pi}}\int\limits_{\frac{\tau}{\sqrt{2}\sigma}}^{\infty} \exp{-t^2}\d t = S_1.
	\end{align}
	The mean volume fraction~$\vfm$ can be also directly related to~$\tau$ by the Gauss error function~$\erf(\cdot)$, i.e.,
	\begin{equation}
		\vfm = 1 - \erf\left(\frac{\tau}{\sqrt{2}\sigma}\right).
	\end{equation}

	Now let us focus on $S_2$.
	We denote the bivariate Gaussian covariance matrix by
	\newcommand{\G}{\tns{\Sigma}}
	\begin{equation}
		\G = \sigma^2\mat{1 & g \\ g & 1}, \quad \text{where }\; g=\cov(\x,\y)/\sigma^2,
	\end{equation}
	where $\cov(\x,\y)$ is the covariance function of the Gaussian field~$\Int(\x;\omega)$.
	Then,
	\begin{equation}
		\det\G = \sigma^4(1-g^2), \qquad \G\inv = \frac{1}{\sigma^2 (1-g^2)}\mat{1 & -g \\ -g & 1}.
	\end{equation}
	Hence, the two-point correlation function of the phase~$\chi$ is given by
	\begin{align}
	S_2(\x,\y) 
	&= \frac{1}{2\pi\sqrt{\det \G}}\int\limits_{\R^2}\chi_{[\abs{\xi_1}>\tau]}\,\chi_{[\abs{\xi_2}>\tau]}\,\exp{-\frac{1}{2}\vct{\xi}\tp\cdot \G\inv\cdot \vct{\xi}}  \d\vct{\xi} \\
	&= \frac{1}{2\pi\sqrt{1-g^2}}\iint_{\left\lbrace(-\infty,-\tau/\sigma]\cup[\tau/\sigma,\infty)\right\rbrace^2}\exp{-\frac{\xi_1^2 - 2g\,\xi_1 \xi_2 + \xi_2^2}{2(1-g^2)}}  \d\xi_1 \d\xi_2 \\
	&= 2a\left(
		\int\limits_{\tau/\sigma}^\infty\int\limits_{\tau/\sigma}^\infty\exp{-\frac{\xi_1^2 - 2g\,\xi_1 \xi_2 + \xi_2^2}{2(1-g^2)}}  \d\xi_1 \d\xi_2 +
		\int\limits_{\tau/\sigma}^\infty\int\limits_{-\infty}^{-\tau/\sigma}\exp{-\frac{\xi_1^2 - 2g\,\xi_1 \xi_2 + \xi_2^2}{2(1-g^2)}}  \d\xi_1 \d\xi_2  
	\right) \\
	&= 2a\int\limits_{\tau/\sigma}^\infty\int\limits_{\tau/\sigma}^\infty\left(
	\exp{-\frac{\xi_1^2 - 2g\,\xi_1 \xi_2 + \xi_2^2}{2(1-g^2)}} + \exp{-\frac{\xi_1^2 + 2g\,\xi_1 \xi_2 + \xi_2^2}{2(1-g^2)}} \right) \d\xi_1 \d\xi_2 \\
	&= 4a\int\limits_{\tau/\sigma}^\infty\int\limits_{\xi_2}^\infty\left(\exp{-\frac{(\xi_1 - g \xi_2)^2 + (1-g^2)\xi_2^2}{2(1-g^2)}} + \exp{-\frac{(\xi_1 + g \xi_2)^2 + (1-g^2)\xi_2^2}{2(1-g^2)}}\right) \d\xi_1 \d\xi_2,
	\end{align}
	where $a = \frac{1}{2\pi\sqrt{1-g^2}}$.
	Let us denote with $S_2^-$ and $S_2^+$ the integrals
	\begin{equation}
		S_2^\pm = \frac{2}{\pi\sqrt{1-g^2}} \int\limits_{\tau/\sigma}^\infty\int\limits_{\xi_2}^\infty \exp{-\frac{(\xi_1 \pm g \xi_2)^2}{2(1-g^2)}-\frac{1}{2}\xi_2^2} \d\xi_1 \d\xi_2.
	\end{equation}
	Thus, $S_2=S_2^- + S_2^+$.
	Let us first consider the integral~$S_2^-$.
	After a change of variable $\hat{\xi}_1 = \frac{\xi_1 - g \xi_2}{\sqrt{1-g^2}}$, it becomes
	\begin{align}
	S_2^-(\x,\y) 
	&= \frac{2}{\pi} \int\limits_{\frac{\tau}{\sigma}}^\infty\int\limits_{\frac{\xi_2 - g \xi_2}{\sqrt{1-g^2}}}^\infty\exp{-\frac{1}{2}(\hat{\xi}_1^2 + \xi_2^2)}  \d\hat{\xi}_1 \d\xi_2 
	= \frac{2}{\pi} \int\limits_{\frac{\tau}{\sigma}}^\infty\int\limits_{\xi_2\sqrt{\frac{1-g}{1+g}}}^\infty\exp{-\frac{1}{2}(\hat{\xi}_1^2 + \xi_2^2)}  \d\hat{\xi}_1 \d\xi_2\\
	&= \frac{2}{\pi}\!\!\!\int\limits_0^{\sqrt{\frac{1-g}{1+g}}}\!\!\!\frac{\d}{\d z}\left(\int\limits_{\frac{\tau}{\sigma}}^\infty\int\limits_{z\xi_2}^\infty\exp{-\frac{1}{2}(\hat{\xi}_1^2 + \xi_2^2)}  \d\hat{\xi}_1 \d\xi_2\right) \d z +
	\frac{2}{\pi} \int\limits_{\frac{\tau}{\sigma}}^\infty\int\limits_{0}^\infty\exp{-\frac{1}{2}(\hat{\xi}_1^2 + \xi_2^2)}  \d\hat{\xi}_1 \d\xi_2\\
	&= -\frac{2}{\pi} \int\limits_0^{\sqrt{\frac{1-g}{1+g}}}\left(\int\limits_{\frac{\tau}{\sigma}}^\infty\exp{-\frac{1}{2}(z^2\xi_2^2 + \xi_2^2)} \xi_2\d\xi_2 \right) \d z + \vfm \\
	&= -\frac{2}{\pi} \int\limits_0^{\sqrt{\frac{1-g}{1+g}}}\left(\int\limits_{-\infty}^{-\frac{1}{2}(\frac{\tau}{\sigma})^2 (z^2+1)}\exp{x} \d x \right) \frac{\d z}{z^2+1} + \vfm \\
	&= \vfm - \frac{2}{\pi} \int\limits_0^{\sqrt{\frac{1-g}{1+g}}}\exp{-\frac{1}{2}(\frac{\tau}{\sigma})^2 (z^2+1)} \frac{\d z}{z^2+1}, \label{eq:apx:S2-_proof}
	\end{align}
	which can be written as
	\begin{equation}
		S_2^-(\x,\y) = \vfm - 4T\left(\frac{\tau}{\sigma}, \sqrt{\frac{1-g}{1+g}} \right),
	\end{equation}
	where
	\begin{equation}
		T(\tau,x)=\frac{1}{2\pi}\int\limits_{0}^{x}\exp{-\frac{\tau^2}{2}(t^2+1)} \frac{\d t}{t^2+1}
	\end{equation}
	is Owen's T function~\cite{owen1956tables,patefield2000fast}.
	Similarly, changing the variable $\hat{\xi}_1 = \frac{\xi_1 + g \xi_2}{\sqrt{1-g^2}}$ in~$S_2^+$ leads to
	\begin{align}
	S_2^+(\x,\y) 
	&= \frac{2}{\pi} \int\limits_{\tau/\sigma}^\infty\int\limits_{\xi_2\sqrt{\frac{1+g}{1-g}}}^\infty\exp{-\frac{1}{2}(\hat{\xi}_1^2 + \xi_2^2)}  \d\hat{\xi}_1 \d\xi_2\\
	&= \vfm - \frac{2}{\pi} \int\limits_0^{\sqrt{\frac{1+g}{1-g}}}\exp{-\frac{1}{2}(\tau/\sigma)^2 (z^2+1)} \frac{\d z}{z^2+1}
	= \vfm - 4T\left(\frac{\tau}{\sigma}, \sqrt{\frac{1+g}{1-g}} \right). \label{eq:apx:S2+_proof}
	\end{align}
	Thus, we have
	\begin{equation}
		S_2(\x,\y) = 2\vfm - 4T\left(\frac{\tau}{\sigma}, \sqrt{\frac{1-g}{1+g}} \right) - 4T\left(\frac{\tau}{\sigma}, \sqrt{\frac{1+g}{1-g}} \right).
	\end{equation}	
	Moreover, from
	\begin{equation}
		T(\tau/\sigma,0) = 0, \quad T(\tau/\sigma,1) = \frac{1}{4}\vfm (1-\frac{1}{2}\vfm), \quad T(\tau/\sigma,\infty) = \frac{1}{4}\vfm,
	\end{equation}
	we have that $S_2(\x,\x)=\vfm$ and $\lim_{\norm{\x-\y}\rightarrow\infty}S_2(\x,\y)=\vfm^2$.
	
	Alternatively, given $T(\tau/\sigma,1) = \frac{1}{4}\vfm (1-\frac{1}{2}\vfm)$,	changing the variable $z = \sqrt{\frac{1-t}{1+t}}$ in~(\ref{eq:apx:S2-_proof}) and $z = \sqrt{\frac{1+t}{1-t}}$ in~(\ref{eq:apx:S2+_proof}) leads respectively to	
	\begin{align}
	S_2^-(\x,\y) 
	&= \vfm - 4T\left(\tau/\sigma, 1 \right) - \frac{2}{\pi} \int\limits_1^{\sqrt{\frac{1-g}{1+g}}}\exp{-\frac{1}{2}(\tau/\sigma)^2 (z^2+1)} \frac{\d z}{z^2+1} \\
	&= \frac{1}{2}\vfm^2 - \frac{2}{\pi} \int\limits_0^{g}\exp{-\frac{(\tau/\sigma)^2}{1+t}}  \; \frac{1}{2}(1+t) \frac{\d}{\d t}\left(\sqrt{\frac{1-t}{1+t}}\right)\d t 
	= \frac{1}{\pi} \int\limits_0^{g}\exp{-\frac{(\tau/\sigma)^2}{1+t}} \frac{\d t}{\sqrt{1-t^2}} + \frac{1}{2}\vfm^2
	\end{align}
	and	
	\begin{align}
	S_2^+(\x,\y) 
	&= \vfm - 4T\left(\tau/\sigma, 1 \right) - \frac{2}{\pi} \int\limits_1^{\sqrt{\frac{1+g}{1-g}}}\exp{-\frac{1}{2}(\tau/\sigma)^2 (z^2+1)} \frac{\d z}{z^2+1} \\
	&= \frac{1}{2}\vfm^2 - \frac{2}{\pi} \int\limits_0^{g}\exp{-\frac{(\tau/\sigma)^2}{1-t}}  \; \frac{1}{2}(1-t) \frac{\d}{\d t}\left(\sqrt{\frac{1+t}{1-t}}\right)\d t 
	= \frac{1}{\pi} \int\limits_0^{g}\exp{-\frac{(\tau/\sigma)^2}{1-t}} \frac{\d t}{\sqrt{1-t^2}} + \frac{1}{2}\vfm^2.
	\end{align}
	Hence,
	\begin{align}
		S_2(\x,\y)
		&= \frac{1}{\pi} \int\limits_0^{g}\left(\exp{-\frac{(\tau/\sigma)^2}{1+t}} + \exp{-\frac{(\tau/\sigma)^2}{1-t}}\right) \frac{\d t}{\sqrt{1-t^2}} + \vfm^2 \\
		&= \frac{1}{\pi} \int\limits_0^{g}\exp{-\frac{(\tau/\sigma)^2}{1-t^2}}\left(\exp{t\frac{(\tau/\sigma)^2}{1-t^2}} + \exp{-t\frac{(\tau/\sigma)^2}{1-t^2}}\right) \frac{\d t}{\sqrt{1-t^2}} + \vfm^2 \\
		&= \frac{2}{\pi} \int\limits_0^{g}\exp{-\frac{(\tau/\sigma)^2}{1-t^2}}\cosh\left((\tau/\sigma)^2\frac{t}{1-t^2}\right) \frac{\d t}{\sqrt{1-t^2}} + \vfm^2.
	\end{align}
	Similar but more general formula can be found in~\cite[eq.(33)]{berk1991scattering}.
\end{proof}

	\bibliographystyle{siamplain}
	\bibliography{myBibliography}

\end{document}